# Machine Learning Aided Computational Discovery of Boride Substrates for Borophene Synthesis

Luqing Wang[1,2,*], Qunfei Zhou[2,6], Qiucheng Li[3], Venkata Surya Chaitanya Kolluru[2,7], Hui Li[3], Albert Tsui[3], Joshua Paul[2], Pierre Darancet[2], Mark C. Hersam[3,4,5], Maria K. Y. Chan[2,8*]

[1]Department of Physics and Astronomy, University of Nevada Las Vegas, Las Vegas, NV 89154, United States
[2]Center for Nanoscale Materials, Argonne National Laboratory, Lemont, IL 60439, United States
[3]Department of Materials Science and Engineering, Northwestern University, Evanston, IL 60208, United States
[4]Department of Chemistry, Northwestern University, Evanston, IL 60208, United States
[5]Department of Electrical and Computer Engineering, Northwestern University, Evanston, IL 60208, United States
[6]Department of Physics and Astronomy, University of Kansas, Lawrence, KS 66045, United States
[7]Department of Materials Science and Engineering, The Ohio State University, Columbus, OH 43210, United States
[8]Northwestern-Argonne Institute for Scientific and Engineering Excellence, Northwestern University, Evanston, IL 60208, United States
* Corresponding authors: luqing.wang@unlv.edu, mchan@anl.gov

**Abstract**

Borophene, a monolayer of boron, is distinctive among two-dimensional materials owing to its unique properties, including polymorphism and potential applications in quantum technologies and energy conversion and storage. Nevertheless, its applications are hindered by the strong binding of borophene to metal substrates, which impedes the realization of freestanding borophene. Alternative substrates with tunable binding to borophene are therefore highly desirable. Since boride layers arise during borophene synthesis on metals, borides provide potential alternative substrates for borophene synthesis. In this work, we investigate the binding energies and interactions between borophene and 1356 boride surfaces for two borophene phases (v1/6 and v1/9), using density functional theory (DFT) calculations and machine learning (ML). Importantly, incorporating ML-predicted charge transfer as an input feature to a subsequent ML model enables physics-informed predictions of binding energies. Additional DFT calculations on four representative cases were performed to validate the ML predictions. SHAP analysis reveals that the majority of key descriptors are associated with delocalized electrons, highlighting the role of electronic properties in governing binding strength. Combined DFT and ML screening identifies 65 and 73 promising boride surfaces for the v1/6 and v1/9 phases, respectively. Al- and Ag-based borides, including $AlB_2$ and $AgB_2$, emerge as predominant candidates for both phases. The computationally predicted candidate substrates and the deduced design guidelines provide a charge-transfer-informed ML strategy for substrate selection and represent promising steps toward the experimental realization of freestanding borophene.

Borophene is special among two dimensional (2D) materials for the lack of a layered bulk counterpart. The rich variety of borophene structures cannot be deduced from bulk boron phases, and is the subject of myriad studies [1-6]. Borophene possesses novel properties in terms of both fundamental science and versatile applications.[7–12] Unlike other 2D materials, it has polymorphism, as various stable phases have been found, such as v1/6, v1/5, v1/9 and so on.[5] Since the structure is deeply intertwined with the properties, polymorphism signifies that diverse properties can be achieved.[8] These characteristics originate from the electron-deficient *p* orbital of boron atoms resulting in a delocalized multi-center bonding (combination of three-center, two-electron and two-center, two-electron bonding ) between boron atoms. Moreover, unlike the bulk phases of boron which are semiconducting, borophene is metallic with massless Dirac fermions and phonon mediated superconductivity as well.[13,14] Besides these exceptional properties, borophene is also attractive for versatile applications for quantum information science [15, 16], and energy conversion [17-20] and storage [21-23]such as for catalysis[25] and batteries[26].

Borophene is a synthetic 2D material. Its structure was originally theoretically predicted prior to experimental synthesis. The early investigation of boron clusters as precursor configurations, such as fullerene-like $B_{80}$[27] and quasi-planar $B_{36}$[28], demonstrated the stability and the potential formation of 2D planar boron structures with hexagonal vacancies. Borophene α phase structure was theoretically predicted in 2007.[2,29] The first experimental realization of borophene was in 2015 on Ag substrate with v1/6 and v1/5 phases.[30,31] So far, borophene has been synthesized on metal substrates, including Ag[30,31], Al[32], Au[33], Cu[34–36], Ir[37–40], and Ru[41]. Nevertheless, there is a major challenge in this field that freestanding borophene cannot be obtained because of the strong binding between borophene and the metal substrates.[7] This impedes the application of borophene in electronic devices. Metal substrates also have other limitations, such as the hindrance of transport measurements and the suppression of superconductivity. Hence, there is a need to find appropriate alternative substrates for borophene synthesis. A theoretical study has suggested step-edge epitaxial borophene growth on hBN.[42] Noteworthily, borides have recently been observed upon deposition of boron atoms onto metal substrates, as the deposited boron atoms induce reconstruction of the topmost metal layer.[43-46] Borides have the potential to be more versatile substrates compared to metals for borophene synthesis. However, there is no guideline for the choice of the appropriate substrates for borophene synthesis yet.

In this work, we search for promising boride substrates for borophene synthesis by computationally investigating the binding between borophene and various boride surfaces. Density functional theory (DFT) calculations are used to determine the binding energies and charge transfer between borophene and substrates. To further expand the chemical space of this search for boride substrate, we train machine learning (ML) models on the results of the DFT calculations, and further evaluate ~1300 substrates to suggest promising boride substrates for borophene synthesis.

## Results

The key to obtaining a suitable substrate for 2D materials growth lies in careful tuning of the 2D-substrate binding strength. An appropriate substrate should have moderate binding with the sample, which is neither too weak nor too strong. If the binding is too weak, the formation of nucleation cores will be difficult, and the sample cannot be synthesized. Whereas if the binding is too strong, the separation of 2D material from the substrate is impeded, and freestanding 2D layer cannot be achieved. This work aims at searching for appropriate boride substrates for borophene growth on account of the binding and interaction between the substrate and borophene. The binding energy is defined as

$$E_{binding} = E_{hetero} - E_{substrate} - E_{borophene} \quad \text{(Eq. 1)}$$

where $E_{hetero}$, $E_{substrate}$, and $E_{borophene}$ are the area-normalized energies of the hetero-structures of the borophene plus substrate, the substrate, and the borophene, respectively.

As the charge transfer (CT) between sample and substrate is commonly considered to play an important role in the binding and interaction, it is informative to explore the correlation between the binding energy and charge transfer. The investigation was divided into three steps, DFT dataset generation, ML models training, and ML prediction, as the flowchart shown in Figure 1. In the DFT dataset generation step, DFT-calculated binding energy ($E_{binding}$) $E_{DFT}$ and charge transfer $CT_{DFT}$ for 188 hetero-structures, substrates with borophene, were calculated with planewave DFT (using VASP). These are the input data for the next step. In the ML models training step, four sets of ML models were built based on the DFT dataset. The first model labeled as Ew/oCT model generates ML-predicted binding energy $E_{ML}$ with features excluding DFT-calculated charge transfer $CT_{DFT}$ while using DFT-calculated binding energy $E_{DFT}$ as the target. The second model labeled as CT model generates ML-predicted charge transfer $CT_{ML}$ and uses DFT-calculated charge transfer $CT_{DFT}$ as the target. The third one labeled as $EwCT_{DFT}$ model and the fourth one labeled as $EwCT_{ML}$ model generate ML-predicted binding energy $E_{ML}$ with features including $CT_{DFT}$, and $CT_{ML}$, respectively, and use $E_{DFT}$ as the target. In the ML prediction step, firstly CT model was applied on a larger dataset with a few thousands hetero-structures to predict the charge transfer $CT'_{ML}$; then $EwCT_{ML}$ model was trained with the ML-predicted $CT_{ML}$ included as a feature, to predict the binding energies $E'_{ML}$ for the thousands of hetero-structures. Finally, the candidate substrates were screened for among thousands of borides based on $E'_{ML}$.

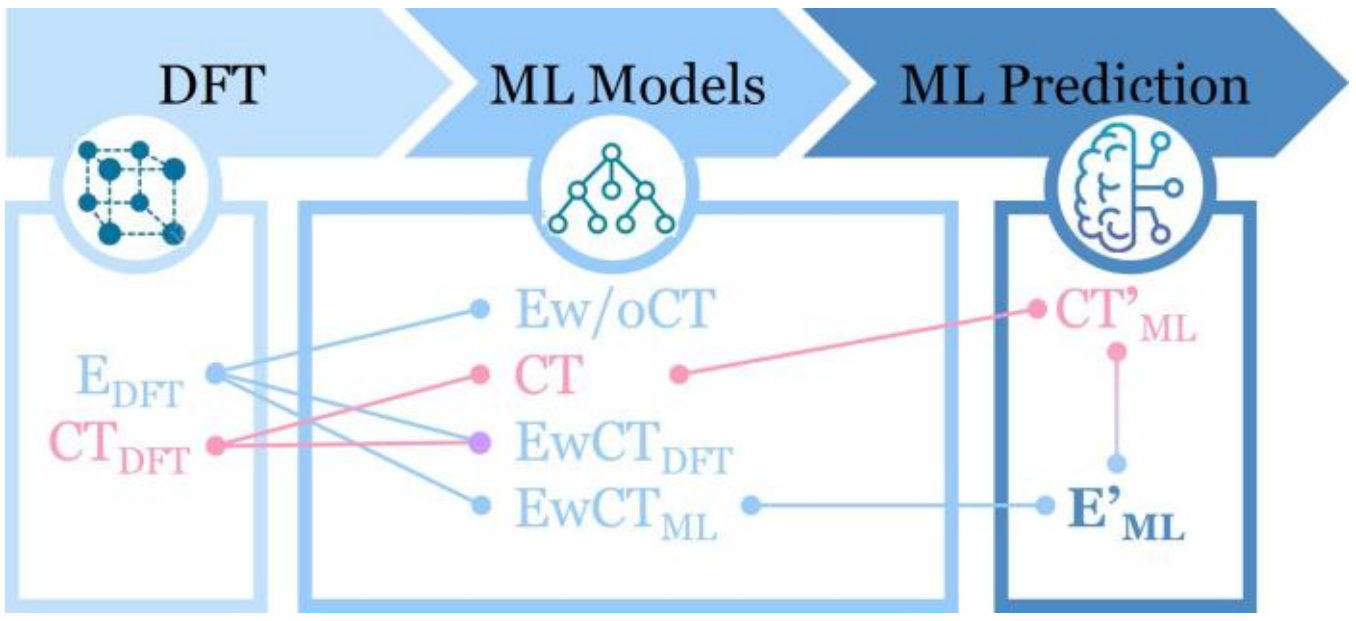

**Figure 1.** The flowchart of the discovery of boride substrates for borophene synthesis.

In the DFT dataset generation step, heterostructures of borophene and boride surfaces were constructed. The boride slabs were cleaved from boride bulk structures and stacked with borophene. To represent the chemical diversity of the borides, 66 binary borides with various crystal structures and chemical formulas including $MB_2$ (M = Be, Mg, Al, Sc, Ti, V, Cr, Mn, Fe, Y, Zr, Nb, Mo, Ru, Ag, Hf, Ta, W, Re, Os, Au, Gd, Lu), MB (M = Ir, Re, Os, Rh, Pt, Zr, Hf, V, Nb, Ta, Cr, Ni, Mo, W, Ti, Fe, Mn, Co), $M_3B_2$ (M = V, Nb, Ta), $M_2B$ (M = Cr, Mo, W, Ta, Mn, Fe, Co, Ni), $M_7B_3$ (M = Re, Tc, Ru, Rh) and $M_3B$ (M = Co, Ni, Pd) were considered.[48] Figure 2 (a-d) shows the typical structures of $MB_2$ and MB. Depending on the crystal symmetry of the boride, each cleaved slab has either two nonequivalent surfaces, one with all boron atoms exposed and the other with all metal atoms exposed, or two identical surfaces which are stoichiometry-balanced with both boron and metal atoms exposed. Here, the three types of surfaces are labeled as B-rich (boron-rich), M-rich (metal-rich), stoichiometry-balanced surfaces, respectively. Examples are shown in Figure 2 (b) and (d). A total of 94 surfaces were generated. Borophene is polymorphic, and a phase is named by the density of vacancies in the triangular lattice.[5] Here, two phases, v1/6 and v1/9 with different symmetries are considered. The v1/6 phase is one of the earliest experimentally synthesized phases (v1/6 and v1/5).[30,31] It is anisotropic with a quasi-linear structure as shown in Figure 2 (e). The v1/9 phase is centrosymmetric and shown in Figure 2 (f). Each heterostructure was constructed by stacking a boride surface with a borophene phase. Various stacking modes were considered by rotation and translation. The one with the minimum lattice mismatch was selected. Figure 2 (g, h) show the top and side views of a heterostructure example. A total of 188 heterostructures were determined by structural relaxation calculations for the two borophene phases.

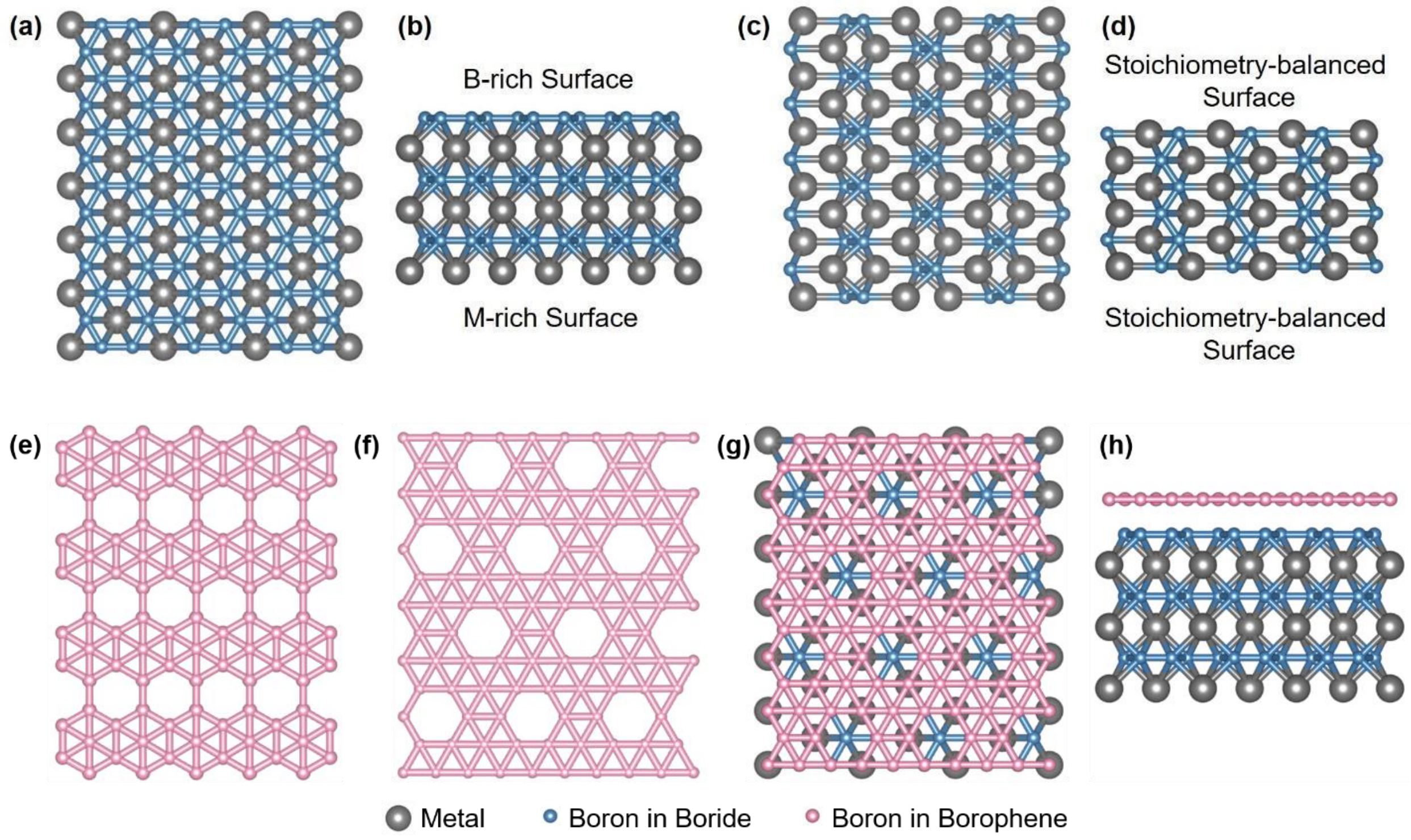

**Figure 2.** The atomic structures. (a) The top and (b) side views of a typical structure of $MB_2$. (c) The top and (d) side views of a typical structure of MB. (e) The v1/6 and (f) v1/9 phases of borophene. (g) The top and (h) side views for a heterostructure of a v1/9 phase on an M-rich surface.

The $E_{binding}$ and the charge transfer of the heterostructures were calculated by performing spin-polarized DFT calculations using the Vienna Ab initio Simulation Package (VASP)[49–52]. The charge transfer was calculated by Bader charge analysis.[53]

Potential substrates for borophene growth can be evaluated based on the value of $E_{binding}$. As discussed earlier, too strong binding results in borophene that cannot be isolated but too weak binding is unfavorable for growth. It has been reported that v1/6 has been synthesized on Ag (111) surface,[30,31] but cannot be separated from the substrate due to too strong binding. Our DFT-calculated $E_{binding}$ for v1/6 on Ag (111) surface is -0.05 eV/Å$^2$. Higher $E_{binding}$ means weaker binding. Thus, the criterion of a potential substrate for borophene growth was set as $E_{binding}$ > -0.05 eV/Å$^2$, meaning that the binding of the heterostructure is weaker than that between v1/6 and Ag (111), for the sake of possible separation. Based on this criterion, 5 boride surfaces were selected for v1/6 growth and 8 for v1/9 by DFT screening. For v1/6 phase growth, the 5 promising substrates are B-rich surfaces of $AlB_2$, $WB_2$, $MnB_2$, $BeB_2$, $MoB_2$. For v1/9 phase growth, the 8 promising substrates are B-rich surfaces of $BeB_2$, $WB_2$, $MoB_2$, $AlB_2$, $AgB_2$, $TaB_2$, and M-rich surfaces of $AlB_2$ and $AgB_2$. Their binding energies are listed in the supplementary information (SI) Table S1.

In order to go beyond the small number of surfaces calculated with DFT, and also to determine the factors that affect binding energies, we train tree-based ML models relating elemental features to DFT binding energies. Feature selection is crucial for ML performance. Here, seven feature sets were constructed for the machine learning models, comprising three individual categories—elemental (E), bulk-averaged (B), and surface-averaged (S)—and four combinations (EB, ES, BS, and EBS). Each set included four basic descriptors: borophene vacancy density, metal-to-boron atomic ratio in the bulk boride, crystallographic space-group number, and surface type (1 for boron-rich, 0 for stoichiometrically balanced, and −1 for metal-rich surfaces). Additionally, 14 elemental properties were considered, either directly representing the metal constituent (E) or compositionally weighted according to the bulk (B) and surface (S) metal-to-boron ratios. The resulting feature sets contained 18, 32, or 46 descriptors, depending on the combination of categories, as summarized in Table 1.

**Table 1.** Feature sets used in machine learning models.

| Feature set | Feature categories | Number of features |
|---|---|---|
| E | Basic + Elemental | 18 |
| B | Basic + Bulk-averaged | 18 |
| S | Basic + Surface-averaged | 18 |
| EB | Basic + Elemental + Bulk-averaged | 32 |
| ES | Basic + Elemental + Surface-averaged | 32 |
| BS | Basic + Bulk-averaged + Surface-averaged | 32 |
| EBS | Basic + Elemental + Bulk-averaged + Surface-averaged | 46 |

**Table note.** The elemental, bulk-averaged, and surface-averaged categories each comprise 14 physicochemical properties: atomic number, atomic weight, empirical radius, electronegativity, electron affinity, ionization energy, melting point, boiling point, density, thermal conductivity, electrical conductivity, specific heat capacity, enthalpy of vaporization, and enthalpy of fusion.

For the regression algorithm, random forest (RF)[54] and XGBoost were applied. RF is an ensemble of decision trees and outputs the mean or average of predictions returned by each tree. XGBoost uses a more regularized model formalization that prunes the trees. To evaluate the ML performance, the coefficient of determination $R^2$ and the root mean square error (RMSE) were adopted as the metrics,[47]

$$R^2 = 1 - \frac{\sum_{i=1}^{N}(Y_i - \hat{Y}_i)^2}{\sum_{i=1}^{N}(Y_i - \bar{Y}_i)^2}$$

$$\mathrm{RMSE} = \sqrt{\frac{1}{N}\sum_{i=1}^{N}(Y_i - \hat{Y}_i)^2}$$

where $Y_i$ are the original labeled data, $\hat{Y}_i$ are the predicted outcomes from the trained ML model, and $\bar{Y}$ is the mean of the original labeled data. The metrics evaluate how much the regression curve fits the original data, determining the performance of the ML models.

ML model training was performed on the DFT dataset. The 188 DFT data were split into training and test sets, with a ratio of 80/20 (150 data in training set and 38 in test). The Ew/oCT model set used the DFT-calculated $E_{binding}$ ($E_{DFT}$) to compute the loss, and used the 2 regression algorithms and 7 feature sets to select the best model. The ML performance metrics and parity plots are listed in Table S2 and SI Figure S1. An ideal model would have test $R^2$ approaching 1, and the test RMSE approaching the training RMSE. For each feature set, XGBoost algorithm has better testing $R^2$ than RF. Feature sets BS and EBS showed the same and the best performance metrics, while set BS includes less features which means it requires less inputs to achieve the same accuracy (performance). Thus, the model using XGBoost algorithm and feature set BS was selected as the best one. Its parity plot is shown in Figure 3 (a).

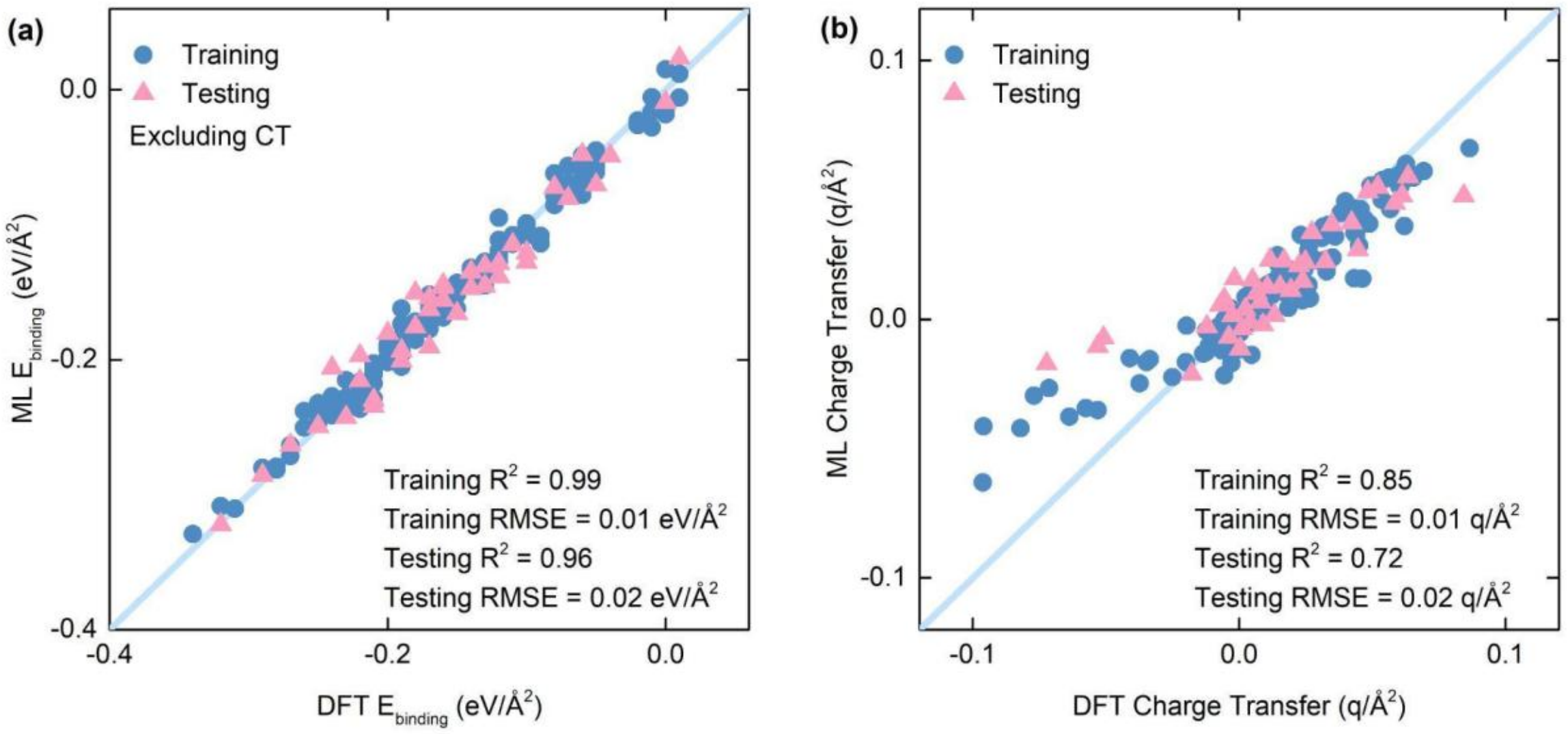

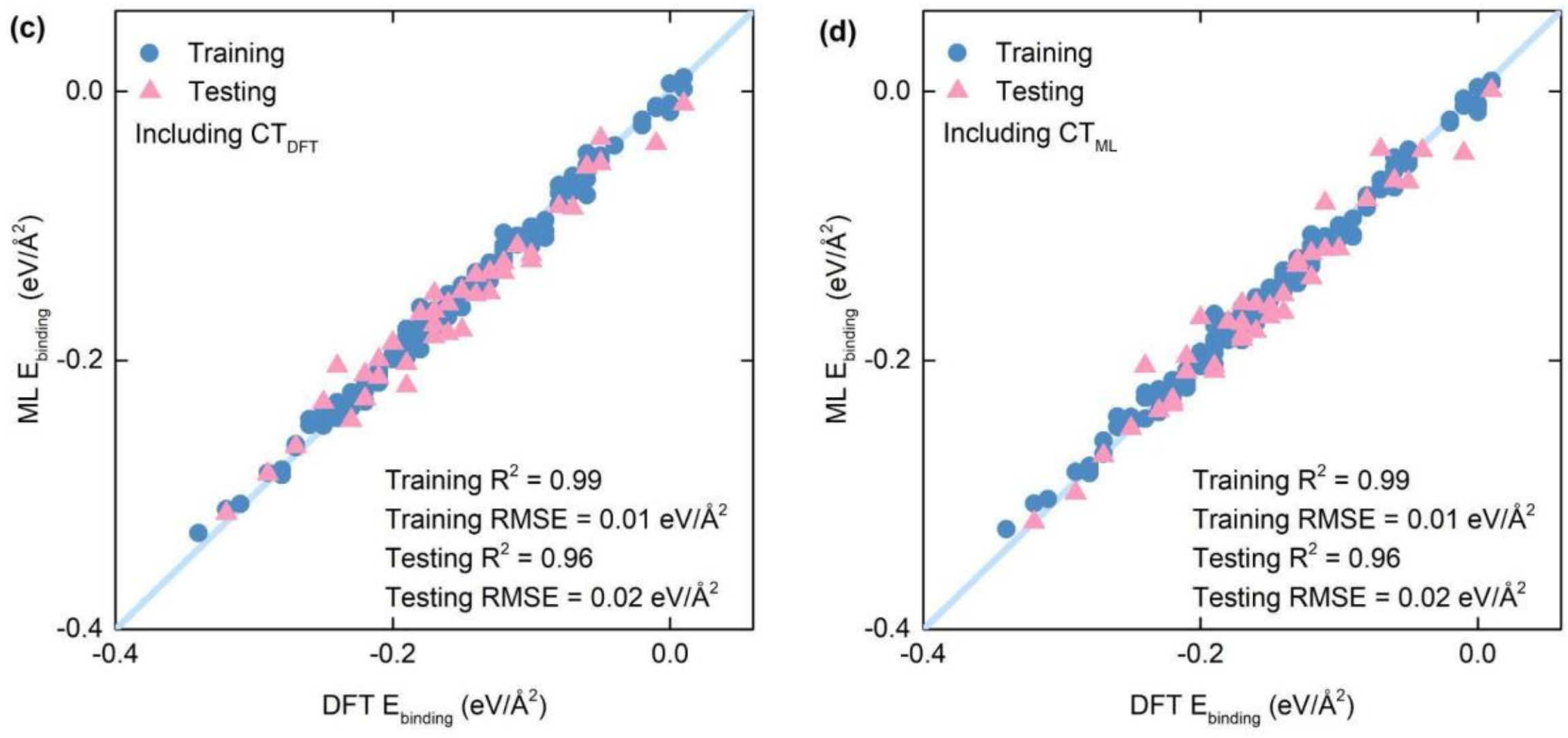
(c)
Training
Testing
Including $CT_{DFT}$
Training $R^2$ = 0.99
Training RMSE = 0.01 eV/Å$^2$
Testing $R^2$ = 0.96
Testing RMSE = 0.02 eV/Å$^2$
ML $E_{binding}$ (eV/Å$^2$)
DFT $E_{binding}$ (eV/Å$^2$)
(d)
Training
Testing
Including $CT_{ML}$
Training $R^2$ = 0.99
Training RMSE = 0.01 eV/Å$^2$
Testing $R^2$ = 0.96
Testing RMSE = 0.02 eV/Å$^2$
ML $E_{binding}$ (eV/Å$^2$)
DFT $E_{binding}$ (eV/Å$^2$)

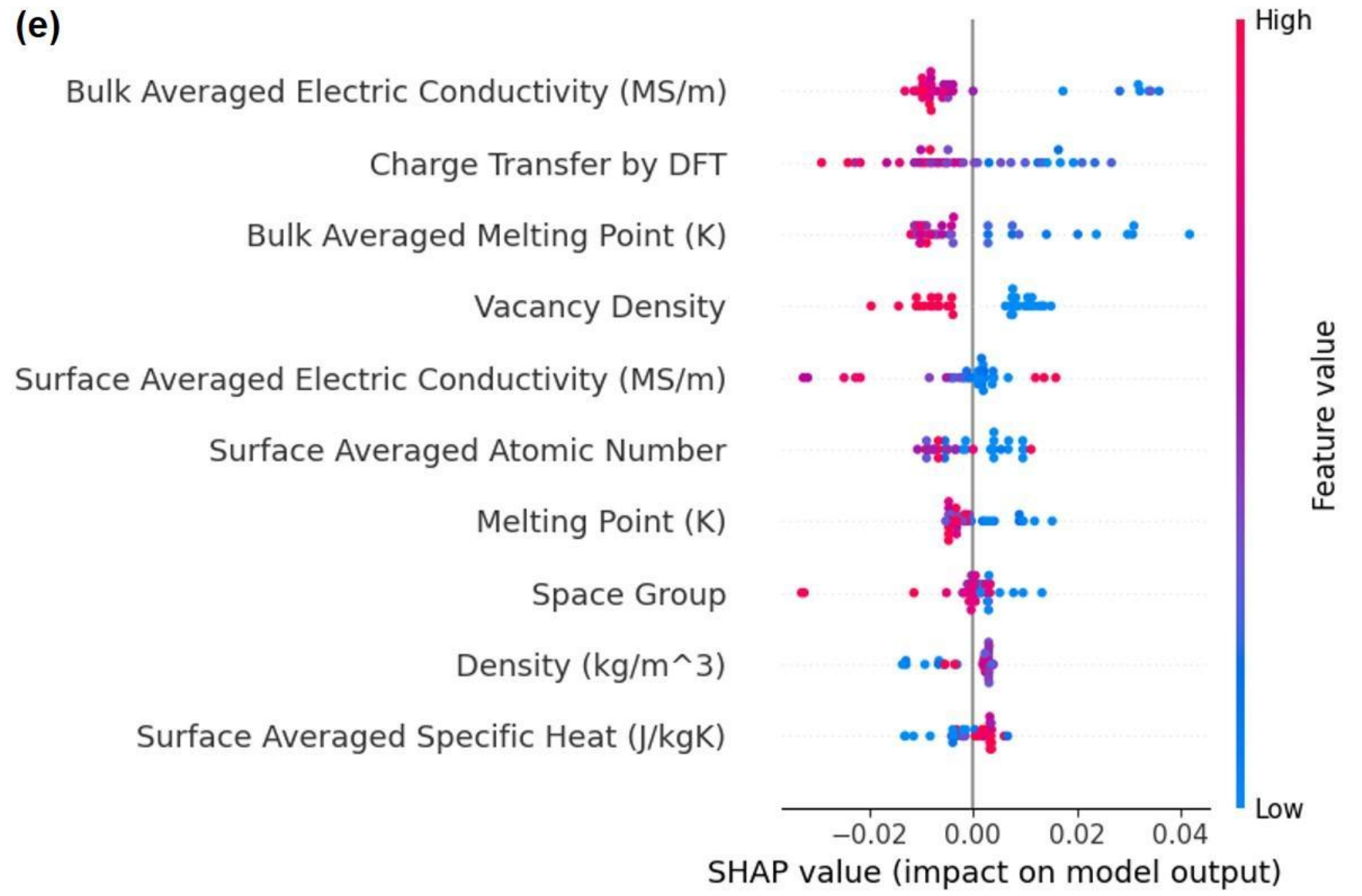
(e)
Bulk Averaged Electric Conductivity (MS/m)
Charge Transfer by DFT
Bulk Averaged Melting Point (K)
Vacancy Density
Surface Averaged Electric Conductivity (MS/m)
Surface Averaged Atomic Number
Melting Point (K)
Space Group
Density (kg/m^3)
Surface Averaged Specific Heat (J/kgK)
SHAP value (impact on model output)
High
Low
Feature value

(f)

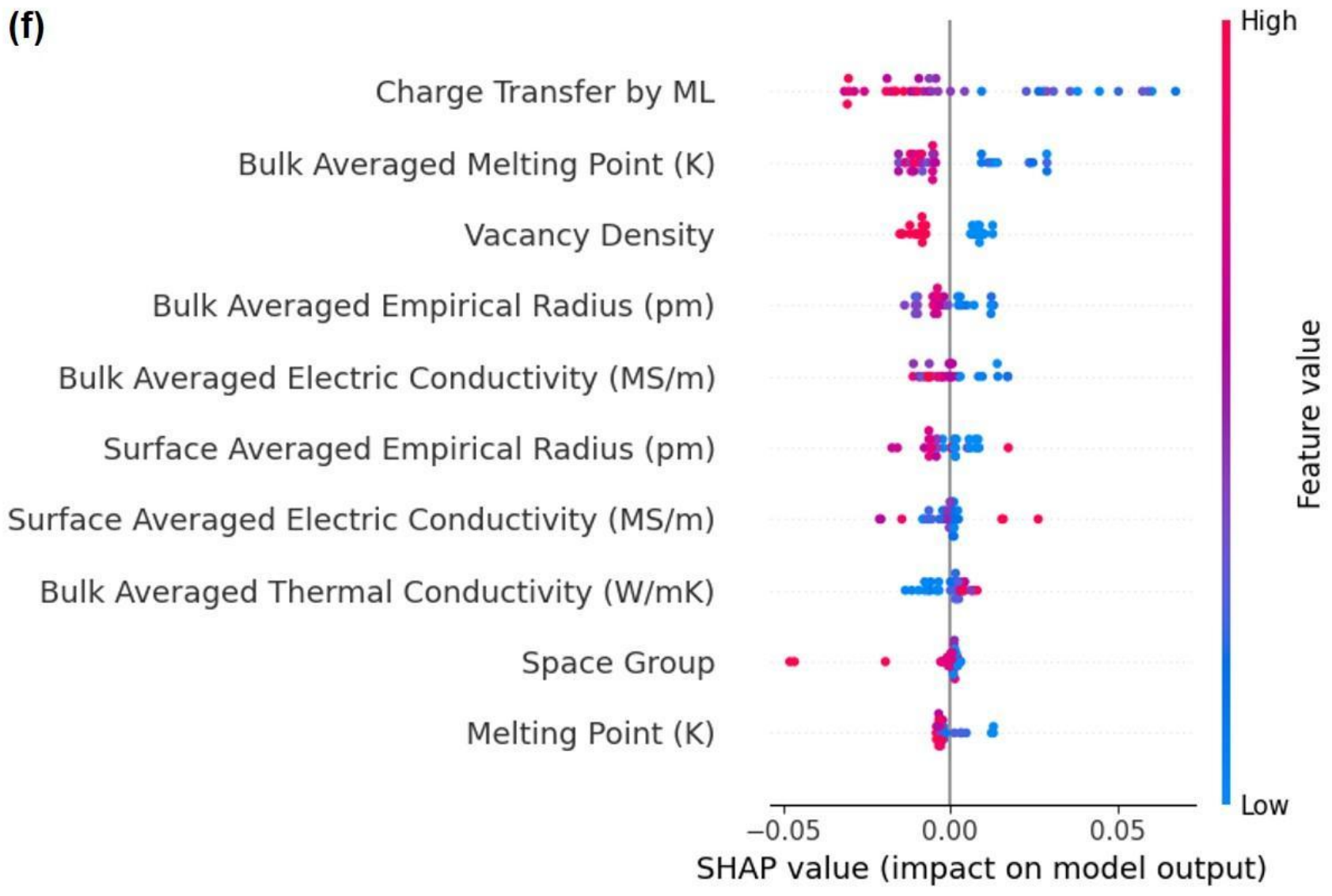

**Figure 3.** ML performance. (a) The Ew/oCT ML model predicting $E_{ML}$. It excludes $CT_{DFT}$ as a feature, and uses XGBoost algorithm and feature set BS. (b) The CT ML model predicting $CT_{ML}$. It uses RF algorithm and feature set EBS. (c) The $EwCT_{DFT}$ ML model predicting $E_{ML}$. It includes $CT_{DFT}$ as one feature, and uses XGBoost and feature set EBS. (d) The $EwCT_{ML}$ ML model predicting $E_{ML}$. It includes $CT_{ML}$ as one feature, and uses XGBoost and feature set EBS. The SHAP summary plots of top 10 features for (e) the $EwCT_{DFT}$ ML model predicting $E_{ML}$ and (f) the $EwCT_{ML}$ ML model predicting $E_{ML}$.

Then the CT model set used the DFT-calculated charge transfer ($CT_{DFT}$) to compute the loss, and used the 2 algorithms and 7 feature sets to select the best model. The ML performance metrics and parity plots are listed in Table S3 and SI Figure S2. Since the model using RF algorithm and feature set EBS has the highest test $R^2$ and best testing RMSE, it was selected as the best model and would be adopted in the following ML prediction step for CT prediction. Its ML parity plot is shown in Figure 3 (b).

Next, the $EwCT_{DFT}$ model set used the DFT-calculated $E_{binding}$ ($E_{DFT}$) to compute the loss, and used the 2 algorithms and 7 feature sets to select the best model. The best model would be a reference for the following fourth model. The ML performance metrics and parity plots are listed in Table S4 and SI Figure S3. For all feature sets, XGBoost models have better testing $R^8$ than RF. The model using the XGBoost and feature set EBS showed the best performance. Its ML parity plot is shown in Figure 3 (c). The charge transfer is commonly regarded as a strongly correlated factor to $E_{binding}$. In order to study the correlation, the contributions of each feature to the $E_{binding}$ prediction were investigated by SHAP (SHapley Additive exPlanations) analysis. A positive SHAP value means the target value increases as the feature value increases, while a negative one means the target value decreases as the feature value increases. The SHAP summary plots of the top 10 most influential features are shown in Figure 3 (e). The charge transfer by DFT calculations was the 2nd most influential feature for predicting $E_{binding}$, confirming the strong correlation between the charge transfer and the binding energy. Thus, it is meaningful to include the charge transfer as a feature to predict $E_{binding}$. Besides, the plot of DFT-calculated binding energies $E_{DFT}$ and charge transfer $CT_{DFT}$ is shown in SI Figure S5. Compared with the Ew/oCT ML model excluding CT, the $EwCT_{DFT}$ ML model including CT has the same ML performance metrics, but it is more meaningful and informative in the view of feature analysis. However, DFT calculations of CT were very time-consuming while ML can generate ML-predicted CT speedily using the CT ML model set, which can be used to substitute the DFT-calculated CT to predict $E_{binding}$. This would significantly accelerate the prediction process of a large dataset which includes thousands of heterostructures or more, since expensive DFT calculations can be avoided. Therefore, we created the $EwCT_{ML}$ model set.

The $EwCT_{ML}$ model set also used $E_{DFT}$ to compute the loss, and used the 2 algorithms and 7 feature sets to select the best feature set for $E_{binding}$ prediction. The ML performance metrics and parity plots are listed in Table S5 and SI Figure S4. For each feature set, XGBoost model has better testing $R^8$ than RF. The model using the XGBoost and feature set EBS showed the best performance. Its ML parity plot is shown in Figure 3 (d). Therefore, the XGBoost and feature set

EBS would be adopted in the following ML prediction step for $E_{binding}$ prediction. The $EwCT_{ML}$ model showed the same ML performance as the $EwCT_{DFT}$ model in ML metrics. It means $CT_{ML}$ performed comparably with $CT_{DFT}$ as a feature to predict $E_{binding}$. Hence, ML can save time for the time-consuming calculations of $CT_{DFT}$. The SHAP summary plots of the top 10 most influential features are shown in Figure 3 (f). The charge transfer, here by ML prediction, is the top 1 influential feature. By comparing the 4 sets of ML models, we found that XGBoost works better for $E_{binding}$ prediction while RF works better for CT prediction in this work.

In the ML prediction step, the trained ML models were applied on a larger dataset which includes 2524 heterostructures (1262 boride surfaces for each borophene phase). The borides include existing structures from Materials Project[55], and hypothetical ones which were created through metal atoms substitutions in the existing structures. Firstly, the CT ML model of RF with feature set EBS in the previous ML models training step was applied on the ML dataset to predict the charge transfer $CT_{ML}$ for the 2524 heterostructures. Then the $EwCT_{ML}$ ML model of XGBoost with feature set EBS in the training step was adopted to predict the $E_{binding}$, and the $CT_{ML}$ generated by the CT model was included as one feature. Based on the ML-predicted $E_{binding}$ and the criterion that $E_{binding}$ is higher than -0.05 eV/Å$^2$ for the possible separation of the borophene and the substrate, 60 boride surfaces were selected for v1/6 growth and 65 for v1/9 by ML screening. The predicted boride substrates by ML screenings and the corresponding $E_{binding}$ are listed in SI Table S6. The candidates by DFT screening and the 10 most promising ones by ML screening for v1/6 and v1/9, and the corresponding $E_{binding}$ are shown in Figure 4.

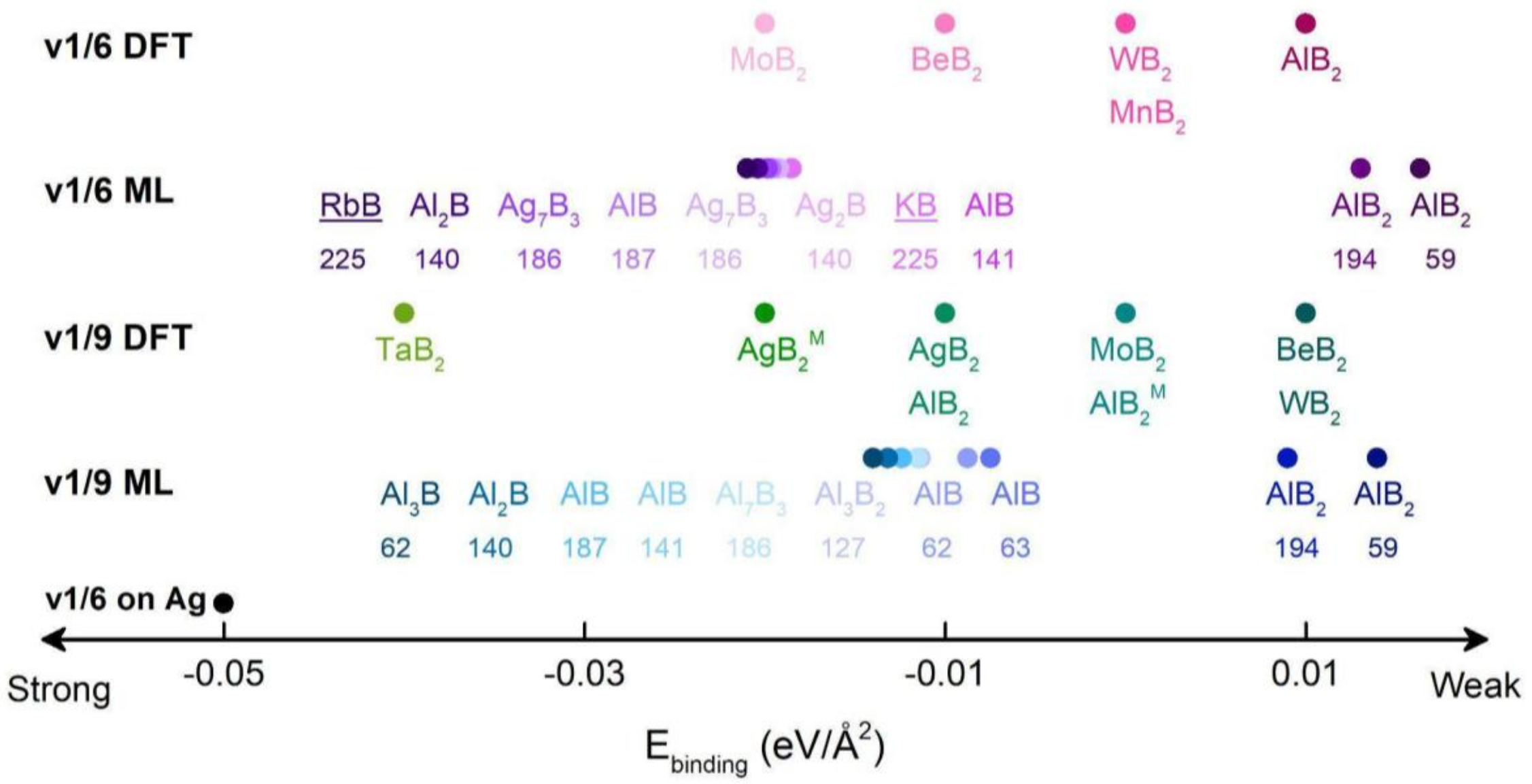

**Figure 4.** The predicted boride substrates for borophene growth by DFT screening, and the 10 most promising substrates identified by ML screening for each borophene phase, and the corresponding $E_{binding}$. The colored points mark the $E_{binding}$. The substrates having the same $E_{binding}$ are in the same color.Underline represents stoichiometry-balanced surface, superscript "M" represents M-rich surface, otherwise it is a B-rich surface. For ML-predicted substrates, the number under a boride is the bulk space group; e.g. "$Al_3B$ 62" represents the B-rich surface of the $Al^3B$ bulk whose space group number is 62. All DFT-predicted boride substrates have space group number 191.

To validate the ML prediction, DFT calculations were performed on some ML-predicted candidates. Since the B-rich surfaces of $AlB_2$ with space group No. 59 and No. 194 are the top 2 ML-predicted candidates for both v1/6 and v1/9 borophene, the two surfaces binding to two borophene phases respectively were selected for DFT validation. Both substrate candidates are hypothetical structures obtained by substituting metal atoms in existing materials. $RuB_2$ (material ID is mp-1077 in Materials Project database) and $ReB_2$ (material ID is mp-1773) are existing materials with space group No. 59 and No. 194 respectively, thus they were taken as parent structure for substitution. The lattice parameters of $RuB_2$ are 2.87 Å, 4.65 Å, and 4.04 Å. After substitution and structural optimization, the lattice parameters of AlB2 with space group No. 59 are 2.93 Å, 3.98 Å, and 4.87 Å. The lattice parameters of $ReB_2$ are 2.91 Å, 2.91 Å, and 7.47 Å, while those of optimized $AlB_2$ with space group No. 194 are 3.00 Å, 3.00 Å, and 7.43 Å. Then slab structures of these two bulks were generated with B-rich surfaces cleaved and optimized. Next, v1/6 and v1/9 were stacked on the two B-rich surfaces respectively. Regarding the stacking configuration, both rotation and shifting were considered, and the heterostructures with the minimum lattice mismatch were selected and relaxed. The optimized four heterostructures are depicted in Figure 5. Based on Eq.1, the DFT-calculated $E_{binding}$ are -0.01 eV/Å$^2$ for v1/9 stacked on B-rich surface of $AlB_2$ with space group No. 194, and 0.00 eV/Å$^2$ for the other three heterostructures. The DFT results demonstrate the weak binding between borophene and candidate substrates, and confirm the ML prediction.

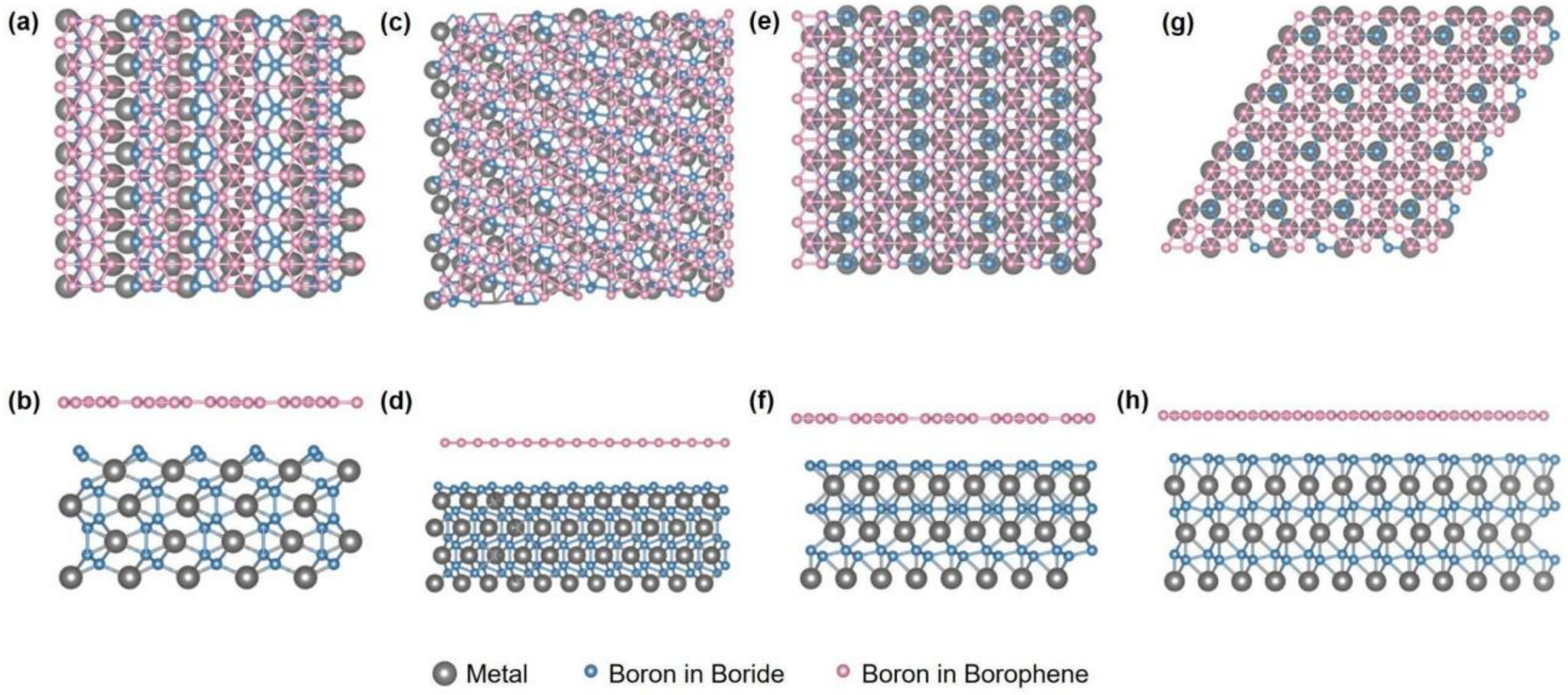

**Figure 5.** Optimized structures by DFT validation. (a) The top and (b) side views of v1/6 on $AlB_2$ with space group No. 59. (c) The top and (d) side views of v1/9 on $AlB_2$ with space group No. 59. (e) The top and (f) side views of v1/6 on $AlB_2$ with space group No. 194. (g) The top and (h) side views of v1/9 on $AlB_2$ with space group No. 194.

## Discussion

Which features are most important? Why are they important and how are they related to binding energy? As shown in Figure 3 (e-f), the most influential features of the ML models are related to electron delocalization (bulk- and surface-averaged electric conductivities, charge transfer, and vacancy density of the borophene phase), with a subset of additional important features related to surface energies (melting point and empirical radius). We interpret those features as reflecting the importance of increasing charge redistribution for binding between borophene and substrate without decreasing interatomic bond strength. Our DFT-screened boride surfaces for both v1/6 and v1/9 phases are $AlB_2$, $WB_2$, $MoB_2$ and $BeB_2$. Besides, two $AgB_2$ surfaces were identified for v1/9. So, the 5 elements Al, W, Mo, Be and Ag are of interest. Regarding melting point, tungsten is the highest among all pure metals. Molybdenum and beryllium are high. Silver is in the mid-range among metals, and aluminum is relatively low. For electric conductivity, silver is the highest of all elements. Aluminum is very high, and beryllium is moderately high. Tungsten and molybdenum are moderate. Since the features used in our ML models were elemental ones weighted by chemical formula, the quantitative relations between features and target were not simply explicit, but the implicit dependencies were incorporated in the ML models which were applied to a larger dataset to screen for more appropriate boride surfaces.

What is the influence of substrate chemical (metal) components for borophene growth? Which metal borides have more candidate surfaces for borophene growth? As shown in Figure 6 , ML screening found 60 surfaces for v1/6 and 65 for v1/9, a significant fraction of which comprised of Al-boride surfaces (17 v1/6 and 19 v1/9) and Ag-boride surfaces (14 and 8). Mg-boride surfaces for v1/6 (6) and Be-boride surfaces for v1/9 (12) are also well represented. This implies that Ag, Al, Be and Mg can be modulated by being combined with boron and forming compounds to have moderate interaction and appropriate affinity with borophene.

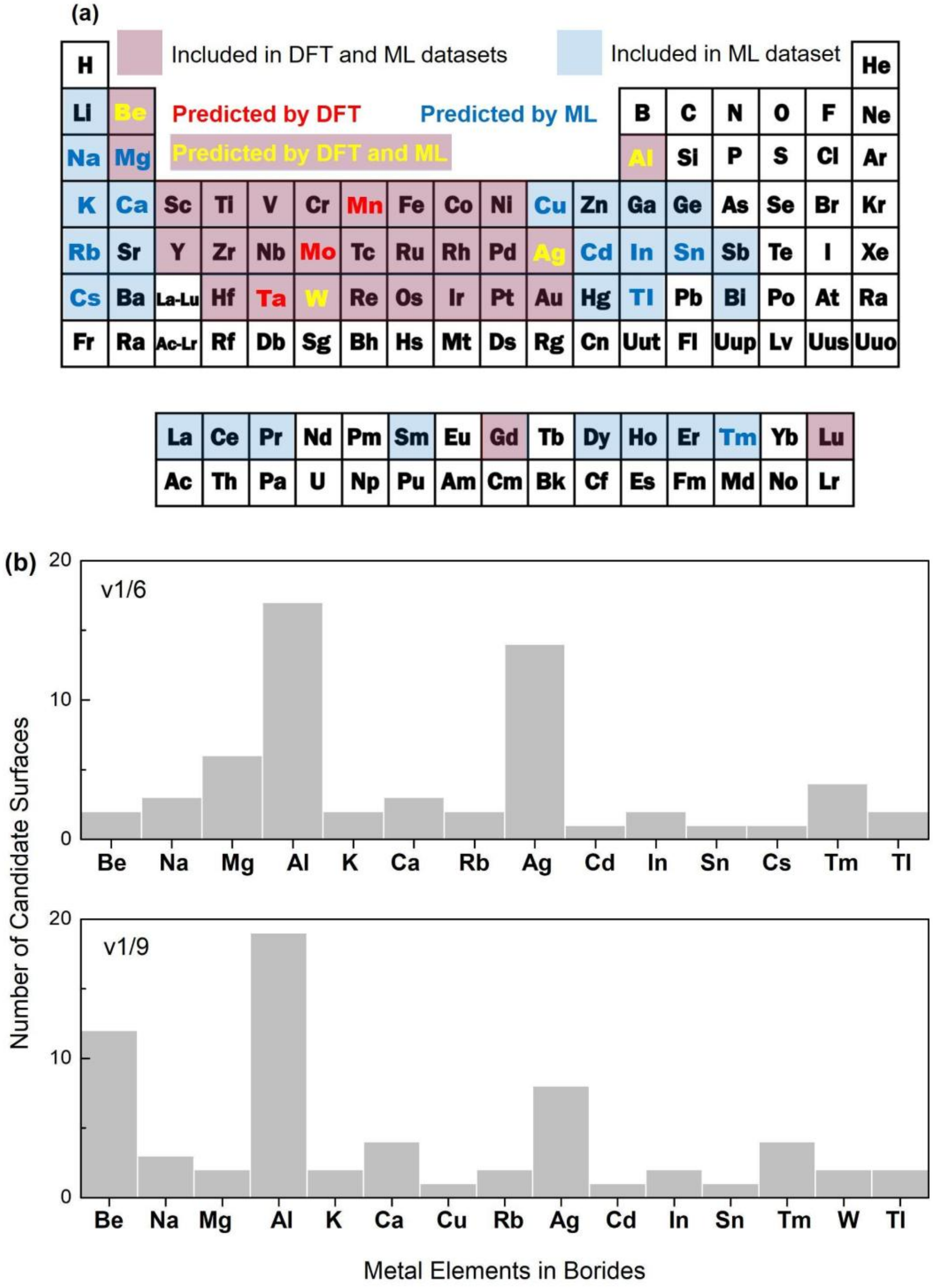

**Figure 6.** Metal elements in predicted borides. (a) Periodic table showing the metal elements included in the DFT and ML datasets and those identified in promising boride substrates. Background shading indicates dataset inclusion, while text colors distinguish elements in DFT- and ML-predicted candidates. (b) Number

of candidate substrate surfaces of the metal borides screened by ML for v1/6 (top) and v1/9 (bottom) synthesis.

Most of the promising substrates are B-rich surfaces. It can be interpreted that the surface boron layer passivates the sub-surface metal layer which is commonly overactive and would have strong binding with borophene, thus the chemical affinity (binding strength) of the surface is modulated to moderate. Besides, Al and Ag have high electric conductivity, therefore they can facilitate large charge transfer from metal to boron within the borides. This leads to less active B-rich surfaces, and therefore smaller binding energy with borophene.

These screened borides also have other novel properties besides being promising as substrates for borophene synthesis. The predominant screened borides are Al-, Ag-, Be- and Mg-borides. $AlB_2$ exhibit superior oxidation resistance compared to binary transition metal diborides.[56] $AgB_2$ is a potential candidate for high-Tc superconductor.[57] $BeB_2$ is a 2D Dirac Material and a promising candidate as an anode material for high-rate sodium-ion batteries.[58] $MgB_2$ is known as a conventional BCS superconductor.[59] Besides, other screened borides also exhibit advanced and novel functionality. For example, $MoB_2$ exhibits high-performance ultraviolet-visible (UV–vis) photodetection.[60] It also can provide high-efficiency hydrogen evolution in both acidic and alkaline environments.[61] Thus, multi-functional novel materials can be expected from these screened borides.

## Conclusions

This work investigated the binding energies and charge transfer between borophene (v1/6 and v1/9 phases) and 1356 boride surfaces, comprising 94 surfaces evaluated by density functional theory (DFT) and 1262 surfaces examined through machine learning (ML) predictions, to identify suitable substrates for borophene synthesis. Combined DFT and ML screening identified 65 and 73 promising boride surfaces for the v1/6 and v1/9 phases, respectively. Al-, Ag-, and Mg-based borides predominate for the v1/6 phase, while Al-, Be-, and Ag-based borides predominate for the v1/9 phase.

A notable aspect of our ML approach is the incorporation of ML-predicted charge transfer as an input feature into a subsequent ML model for predicting borophene–substrate binding energies. This approach integrates physically meaningful information about interfacial electronic interactions into the ML framework, enabling physics-informed predictions and efficient screening of candidate substrates. The ML predictions were further evaluated through additional DFT calculations on four representative candidate substrates. SHAP analysis identifies charge transfer, electrical conductivity, borophene vacancy density, melting point, and empirical radius as influential descriptors of borophene–substrate binding energies. Most of these descriptors are associated with delocalized electrons, highlighting the importance of electronic properties in determining binding strength.

The predominance of Al- and Ag-based borides among the screened candidates suggests that their electronic properties and interactions with boron favor moderate binding to borophene. Moreover, most promising candidates possess boron-rich surfaces, suggesting that surface boron layers passivate reactive subsurface metal atoms and thereby moderate borophene–substrate interactions. Beyond their potential as substrates for borophene synthesis, several identified borides exhibit attractive functional properties, including superconductivity, oxidation resistance, and electrochemical and optoelectronic performance, suggesting opportunities for multifunctional borophene-based heterostructures. Overall, this work provides a physics-informed, ML-assisted screening strategy and practical guidelines for selecting suitable substrates toward the experimental realization of freestanding borophene.

## Methods

DFT calculations were used the Vienna Ab initio Simulation Package (VASP)[49–51]. The structural optimizations and the self-consistent total energy calculations were carried out using the generalized gradient approximation (GGA) and the Perdew−Burke−Ernzerhof (PBE) exchange-correlation functional along with projector-augmented wave (PAW)[52] potentials. Electronic wave functions were expanded in a plane wave basis set with a kinetic energy cutoff of 400 eV. For the Brillouin zone integration, the k-point density is approximately 33 $Å^{-1}$. The energy convergence criterion for the electronic wave function was set to be $10^{-5}$ eV. A vacuum layer of approximately 10 Å (periodic length in the z-direction of 25 ~ 30 Å) was chosen to minimize spurious interactions between the periodic layers.

## Acknowledgement

The computational work was supported by the National Science Foundation Materials Research Science and Engineering Center at Northwestern University under grant number DMR-2308691. Experimental insights were supported by the U.S. Department of Energy under grant number DE-SC0026329-0001. Use of the Center for Nanoscale Materials, a U.S. Department of Energy Office of Science User Facility, was supported by the U.S. DOE, Office of Basic Energy Sciences, under Contract No. DE-AC02-06CH11357. We gratefully acknowledge the computing resources provided by the Laboratory Computing Resource Center at Argonne National Laboratory.

# Supplementary Information

## 1. DFT screening.

**Table S1.** Potential substrates for borophene growth by DFT screening.

| v1/6 Growth | | | | v1/9 Growth | | | |
|---|---|---|---|---|---|---|---|
| No. | Boride | Surface | $E_{binding}$ (eV/Å$^2$) | No. | Boride | Surface | $E_{binding}$ (eV/Å$^2$) |
| 1 | $AlB_2$ | B-rich | 0.01 | 1 | $BeB_2$ | B-rich | 0.01 |
| 2 | $WB_2$ | B-rich | 0.00 | 2 | $WB_2$ | B-rich | 0.01 |
| 3 | $MnB_2$ | B-rich | 0.00 | 3 | $MoB_2$ | B-rich | 0.00 |
| 4 | $BeB_2$ | B-rich | -0.01 | 4 | $AlB_2$ | B-rich | 0.00 |
| 5 | $MoB_2$ | B-rich | -0.02 | 5 | $AgB_2$ | B-rich | -0.01 |
| | | | | 6 | $AlB_2$ | M-rich | -0.01 |
| | | | | 7 | $AgB_2$ | M-rich | -0.02 |
| | | | | 8 | $TaB_2$ | B-rich | -0.04 |

## 2. ML elemental features (14):

atomic number, atomic weight (u), empirical radius (pm), electronegativity, electron affinity (eV), ionization (eV), melting point (K), boiling point (K), density (kg/m3), thermal conductivity (W/mK), electric conductivity (MS/m), specific heat (J/kgK), heat of vaporization (eV), heat of fusion (eV).

## 3. ML performance.

**Table S2.** ML evaluations for the 1st ML model set generating ML-predicted $E_{ML}$ without $CT_{DFT}$.

| | | Feature set | | | | | | |
|---|---|---|---|---|---|---|---|---|
| | | E | B | S | EB | ES | **BS** | EBS |
| RF | Training $R^2$ | 0.95 | 0.95 | 0.95 | 0.95 | 0.95 | 0.95 | 0.95 |
| | Testing $R^2$ | 0.89 | 0.90 | 0.81 | 0.90 | 0.90 | 0.90 | 0.90 |
| | Training RMSE (eV/Å²) | 0.02 | 0.02 | 0.02 | 0.02 | 0.02 | 0.02 | 0.02 |
| | Testing RMSE (eV/Å²) | 0.03 | 0.02 | 0.03 | 0.02 | 0.02 | 0.02 | 0.02 |
| XGBoost | Training $R^2$ | 0.97 | 0.98 | 0.96 | 0.99 | 0.98 | **0.99** | 0.99 |
| | Testing $R^2$ | 0.89 | 0.94 | 0.85 | 0.94 | 0.91 | **0.96** | 0.96 |
| | Training RMSE (eV/Å²) | 0.01 | 0.01 | 0.02 | 0.01 | 0.01 | **0.01** | 0.01 |
| | Testing RMSE (eV/Å²) | 0.02 | 0.02 | 0.03 | 0.02 | 0.02 | **0.02** | 0.02 |

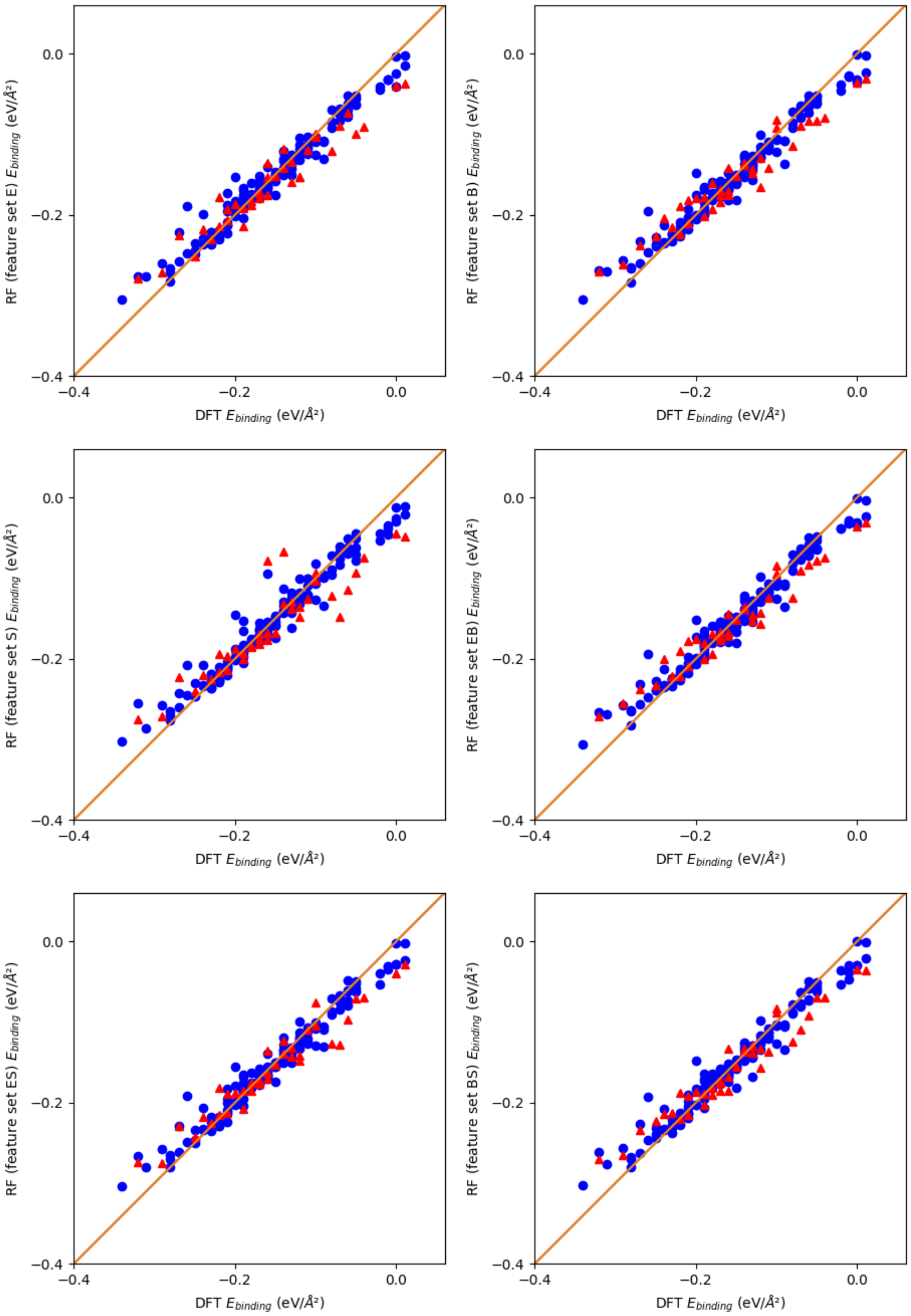

RF (feature set E) $E_{binding}$ (eV/Å²)
DFT $E_{binding}$ (eV/Å²)
RF (feature set B) $E_{binding}$ (eV/Å²)
DFT $E_{binding}$ (eV/Å²)
RF (feature set S) $E_{binding}$ (eV/Å²)
DFT $E_{binding}$ (eV/Å²)
RF (feature set EB) $E_{binding}$ (eV/Å²)
DFT $E_{binding}$ (eV/Å²)
RF (feature set ES) $E_{binding}$ (eV/Å²)
DFT $E_{binding}$ (eV/Å²)
RF (feature set BS) $E_{binding}$ (eV/Å²)
DFT $E_{binding}$ (eV/Å²)
0.0
−0.2
−0.4

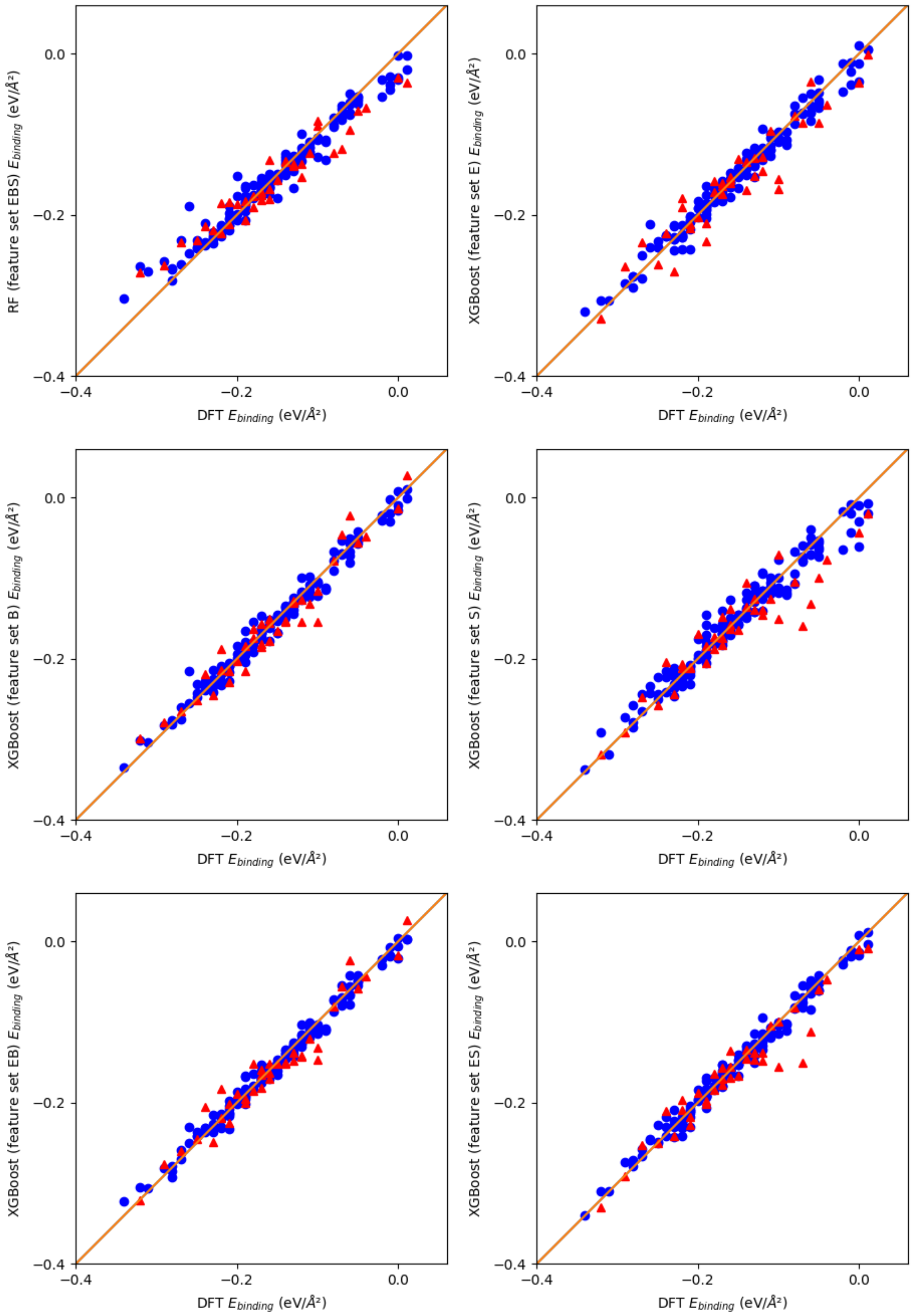
RF (feature set EBS) $E_{binding}$ (eV/Å²)
XGBoost (feature set E) $E_{binding}$ (eV/Å²)
XGBoost (feature set B) $E_{binding}$ (eV/Å²)
XGBoost (feature set S) $E_{binding}$ (eV/Å²)
XGBoost (feature set EB) $E_{binding}$ (eV/Å²)
XGBoost (feature set ES) $E_{binding}$ (eV/Å²)
DFT $E_{binding}$ (eV/Å²)
−0.4
−0.2
0.0

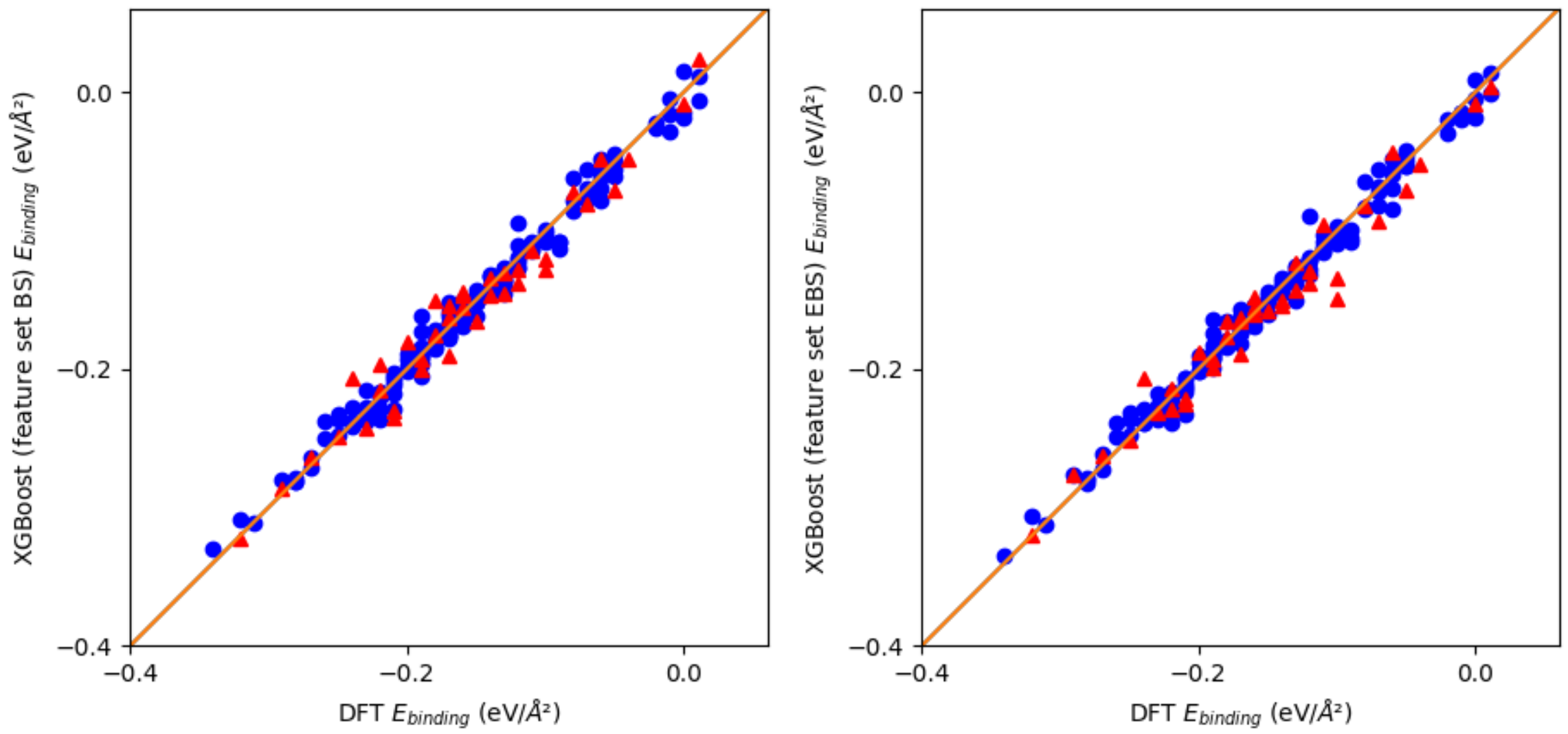

**Figure S1.** ML fitting plots for the 1st ML model set fitting $E_{DFT}$ without $CT_{DFT}$.

**Table S3.** ML evaluations for the 2nd ML model set generating ML-predicted $CT_{ML}$.

| | | Feature set | | | | | | |
|---|---|---|---|---|---|---|---|---|
| | | E | B | S | EB | ES | BS | **EBS** |
| RF | Training $R^2$ | 0.86 | 0.86 | 0.85 | 0.86 | 0.86 | 0.86 | **0.85** |
| | Testing $R^2$ | 0.65 | 0.71 | 0.60 | 0.71 | 0.65 | 0.69 | **0.72** |
| | Training RMSE (q/Å²) | 0.01 | 0.01 | 0.01 | 0.01 | 0.01 | 0.01 | **0.01** |
| | Testing RMSE (q/Å²) | 0.02 | 0.02 | 0.02 | 0.02 | 0.02 | 0.02 | **0.02** |
| XGBoost | Training $R^2$ | 0.88 | 0.92 | 0.89 | 0.91 | 0.89 | 0.91 | 0.92 |
| | Testing $R^2$ | 0.38 | 0.59 | 0.64 | 0.54 | 0.62 | 0.64 | 0.59 |
| | Training RMSE (q/Å²) | 0.01 | 0.01 | 0.01 | 0.01 | 0.01 | 0.01 | 0.01 |
| | Testing RMSE (q/Å²) | 0.03 | 0.02 | 0.02 | 0.02 | 0.02 | 0.02 | 0.02 |

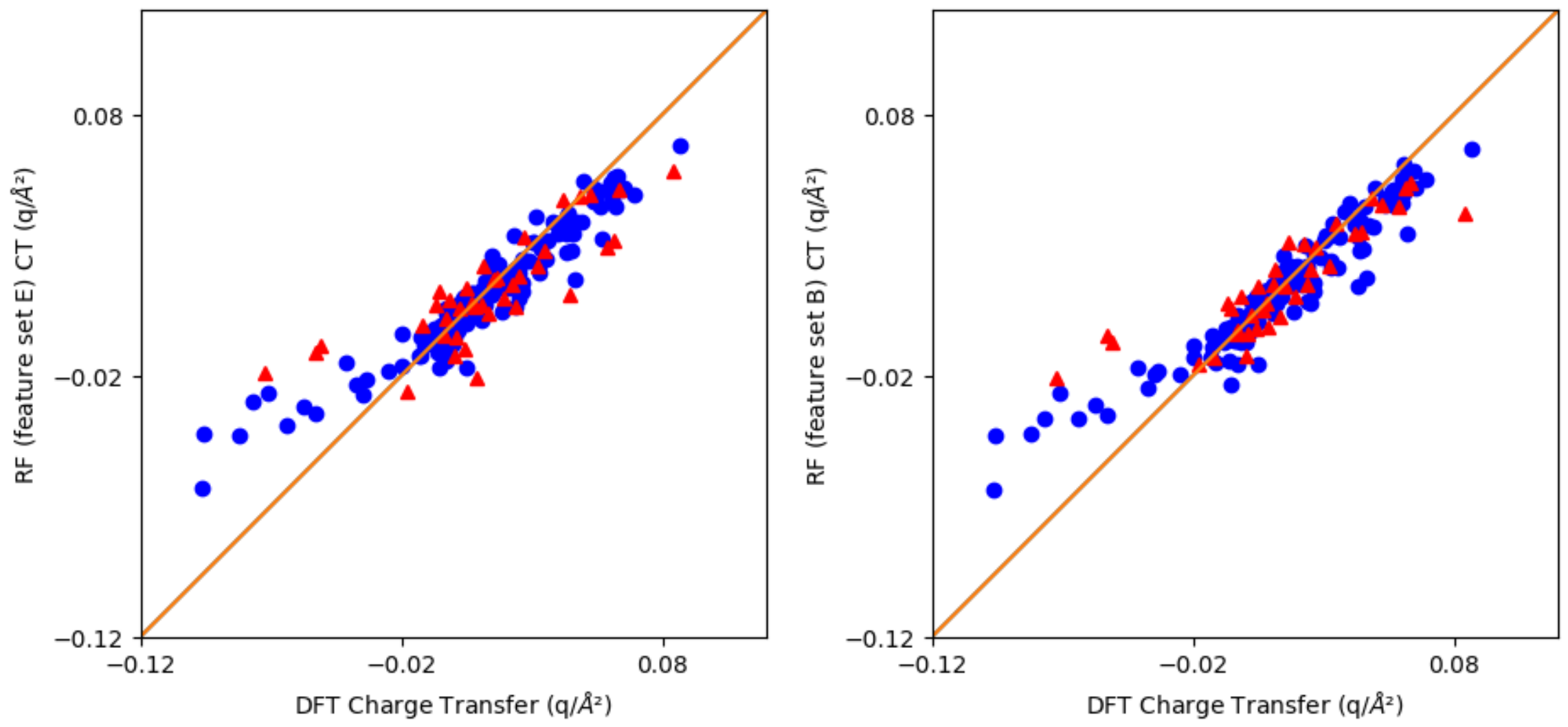

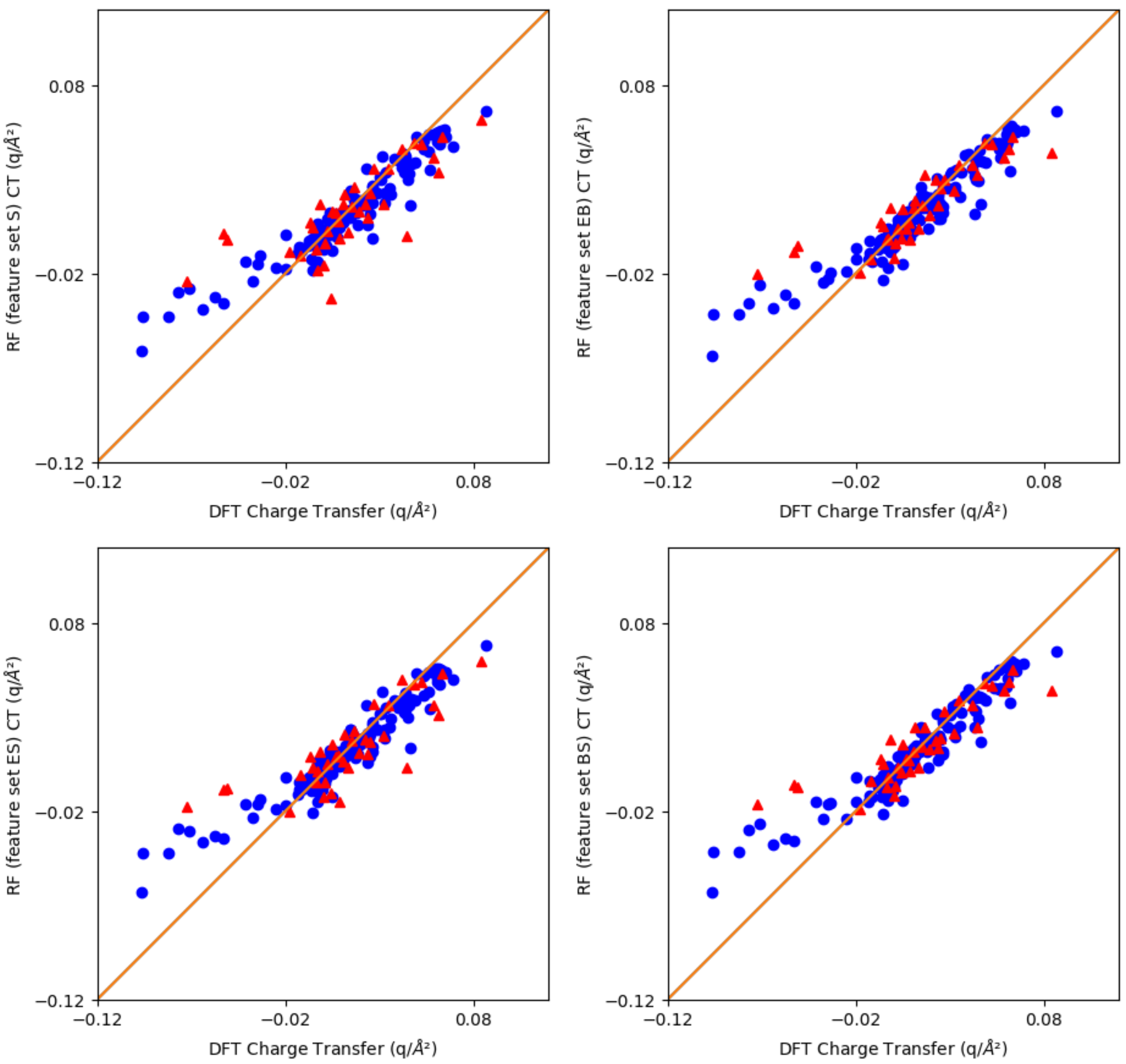
0.08
−0.02
−0.12
−0.12
−0.02
0.08
RF (feature set S) CT (q/Å²)
DFT Charge Transfer (q/Å²)
RF (feature set EB) CT (q/Å²)
RF (feature set ES) CT (q/Å²)
RF (feature set BS) CT (q/Å²)

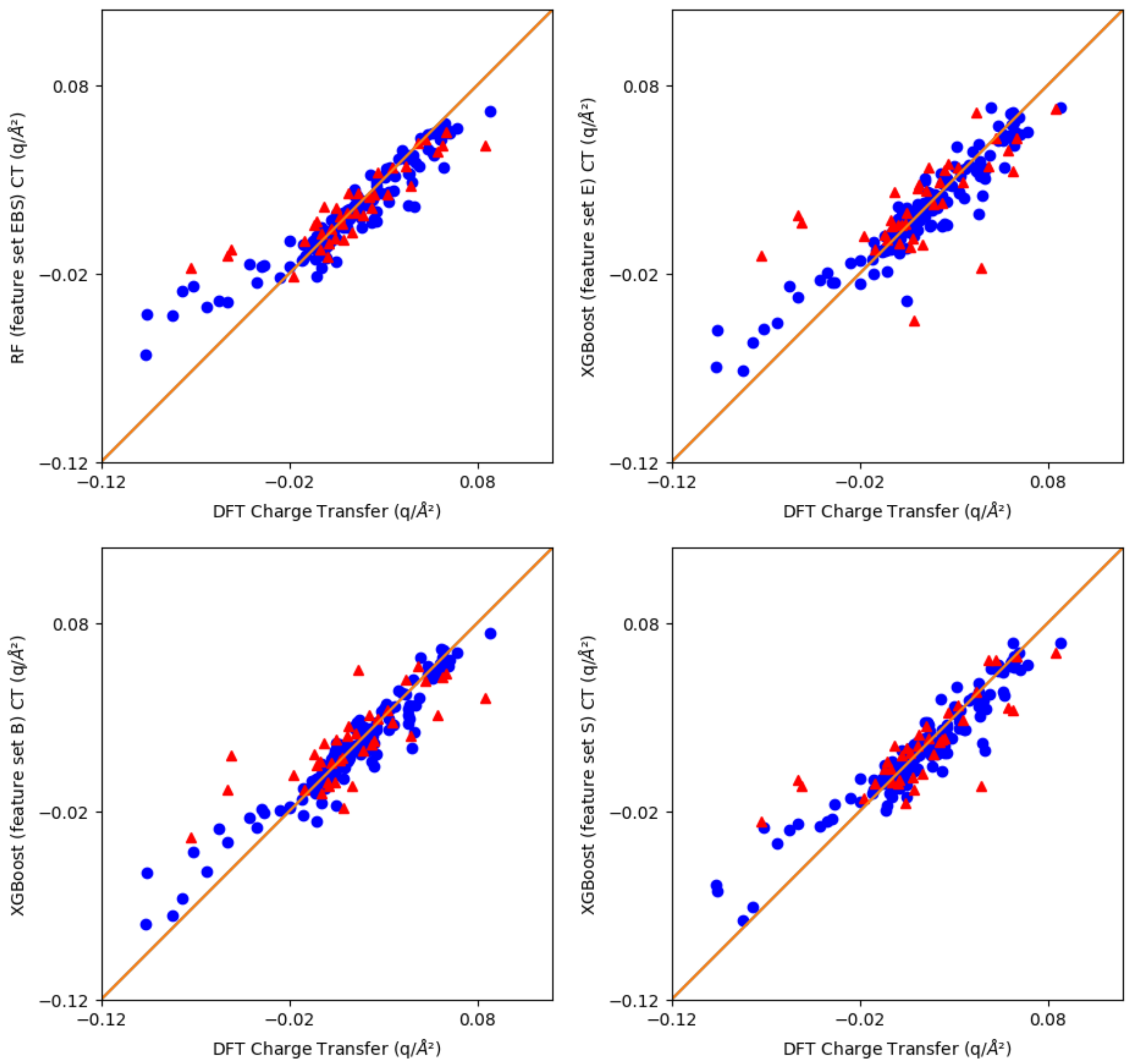
0.08
−0.02
−0.12
RF (feature set EBS) CT (q/Å²)
−0.12
−0.02
0.08
DFT Charge Transfer (q/Å²)
XGBoost (feature set E) CT (q/Å²)
XGBoost (feature set B) CT (q/Å²)
XGBoost (feature set S) CT (q/Å²)

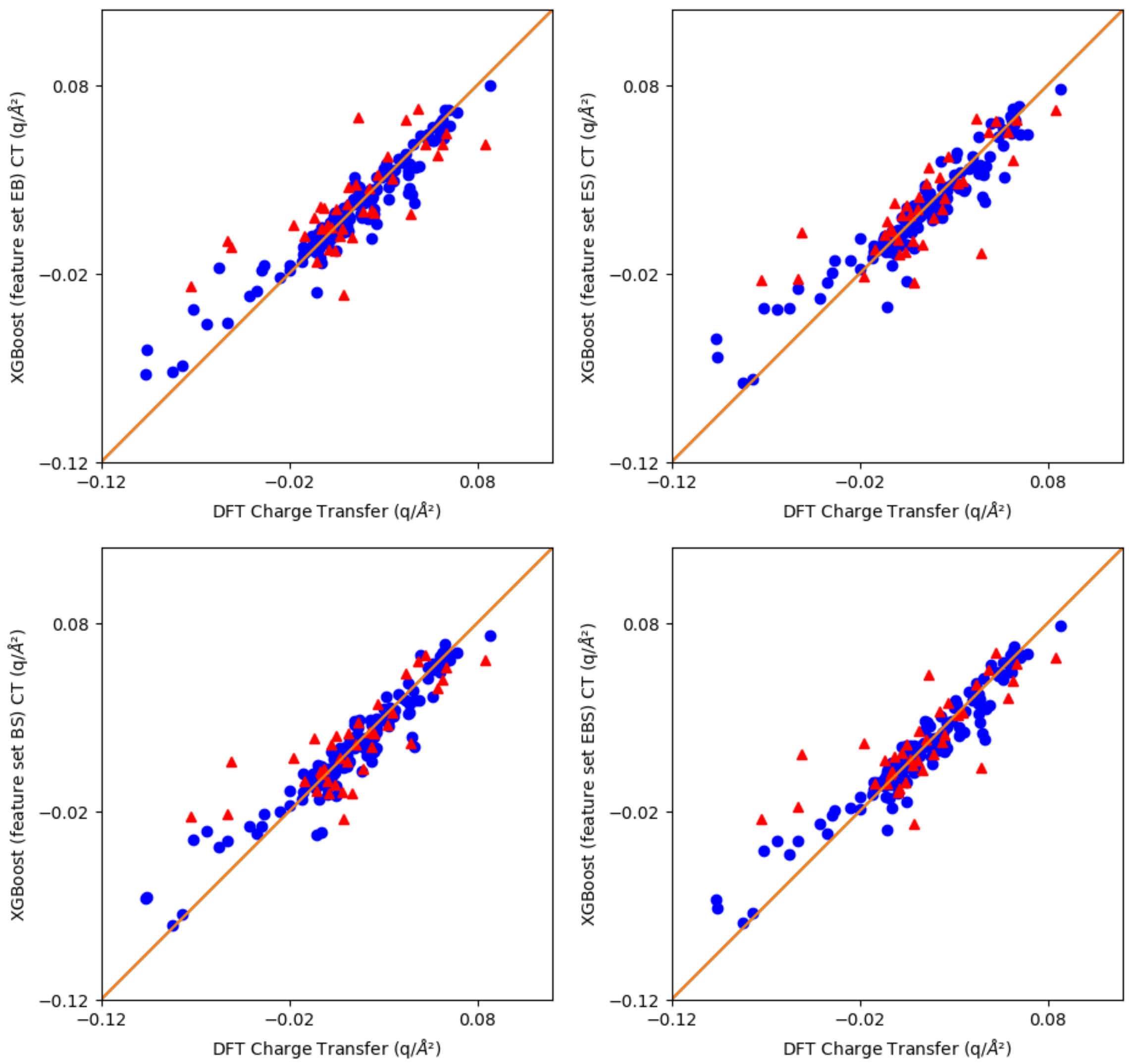

**Figure S2.** ML fitting plots for the 2nd ML model set fitting $CT_{DFT}$.

**Table S4.** ML evaluations for the 3rd ML model set generating ML-predicted $E_{ML}$ including $CT_{DFT}$ as one feature.

| | | Feature set | | | | | | |
|---|---|---|---|---|---|---|---|---|
| | | E | B | S | EB | ES | BS | **EBS** |
| RF | Training $R^2$ | 0.94 | 0.94 | 0.93 | 0.94 | 0.95 | 0.94 | 0.95 |
| | Testing $R^2$ | 0.83 | 0.86 | 0.85 | 0.85 | 0.88 | 0.86 | 0.87 |
| | Training RMSE (eV/Å²) | 0.02 | 0.02 | 0.02 | 0.02 | 0.02 | 0.02 | 0.02 |
| | Testing RMSE (eV/Å²) | 0.03 | 0.03 | 0.03 | 0.03 | 0.03 | 0.03 | 0.03 |
| XGBoost | Training $R^2$ | 0.99 | 0.99 | 0.97 | 0.99 | 0.99 | 1.00 | **0.99** |
| | Testing $R^2$ | 0.91 | 0.89 | 0.92 | 0.89 | 0.94 | 0.93 | **0.96** |
| | Training RMSE (eV/Å²) | 0.01 | 0.01 | 0.01 | 0.01 | 0.01 | 0.01 | **0.01** |
| | Testing RMSE (eV/Å²) | 0.02 | 0.03 | 0.02 | 0.02 | 0.02 | 0.02 | **0.02** |

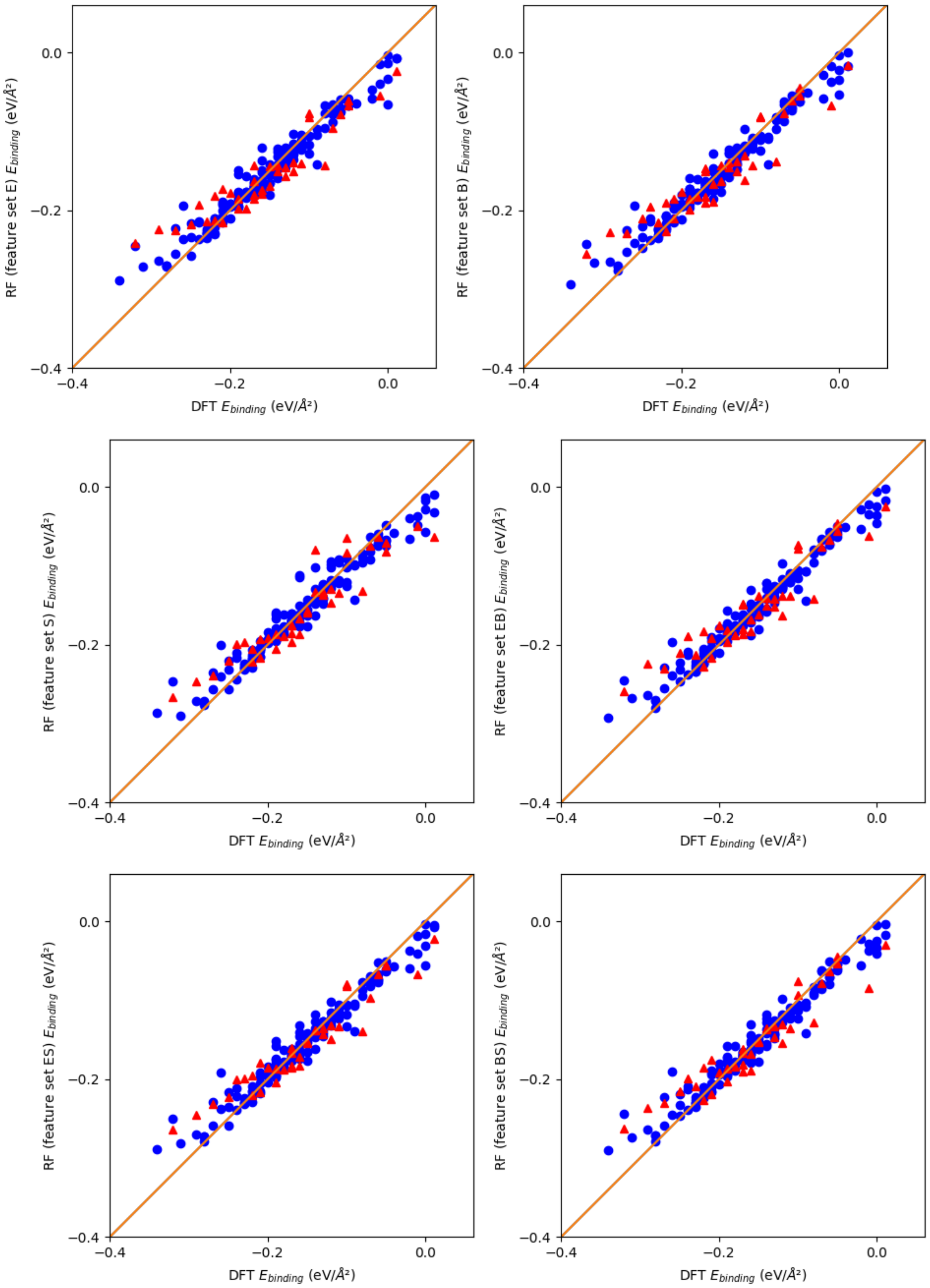
RF (feature set E) $E_{binding}$ (eV/Å²)
RF (feature set B) $E_{binding}$ (eV/Å²)
RF (feature set S) $E_{binding}$ (eV/Å²)
RF (feature set EB) $E_{binding}$ (eV/Å²)
RF (feature set ES) $E_{binding}$ (eV/Å²)
RF (feature set BS) $E_{binding}$ (eV/Å²)
DFT $E_{binding}$ (eV/Å²)
0.0
−0.2
−0.4

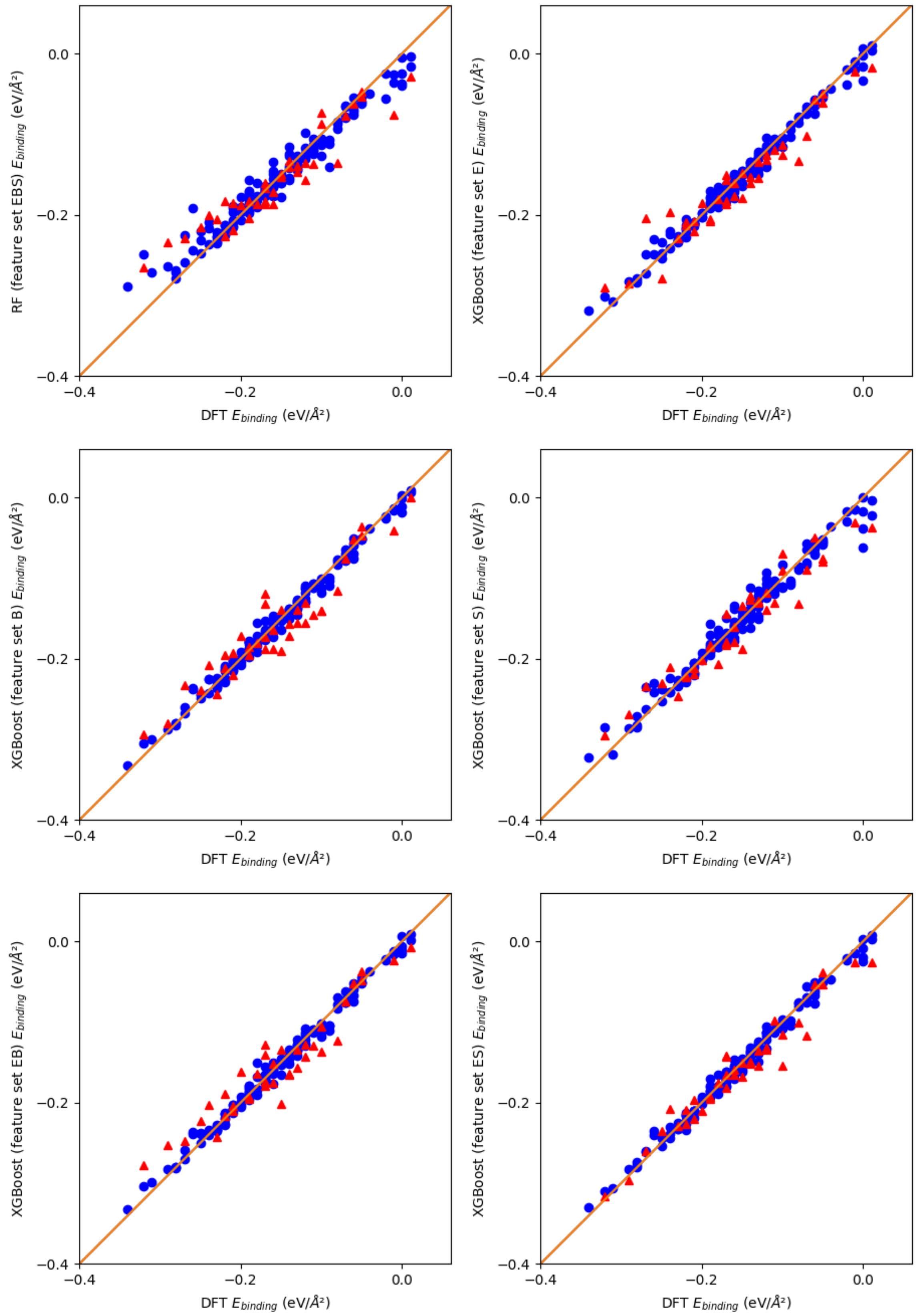
RF (feature set EBS) $E_{binding}$ (eV/Å²)
XGBoost (feature set E) $E_{binding}$ (eV/Å²)
XGBoost (feature set B) $E_{binding}$ (eV/Å²)
XGBoost (feature set S) $E_{binding}$ (eV/Å²)
XGBoost (feature set EB) $E_{binding}$ (eV/Å²)
XGBoost (feature set ES) $E_{binding}$ (eV/Å²)
DFT $E_{binding}$ (eV/Å²)
0.0
−0.2
−0.4

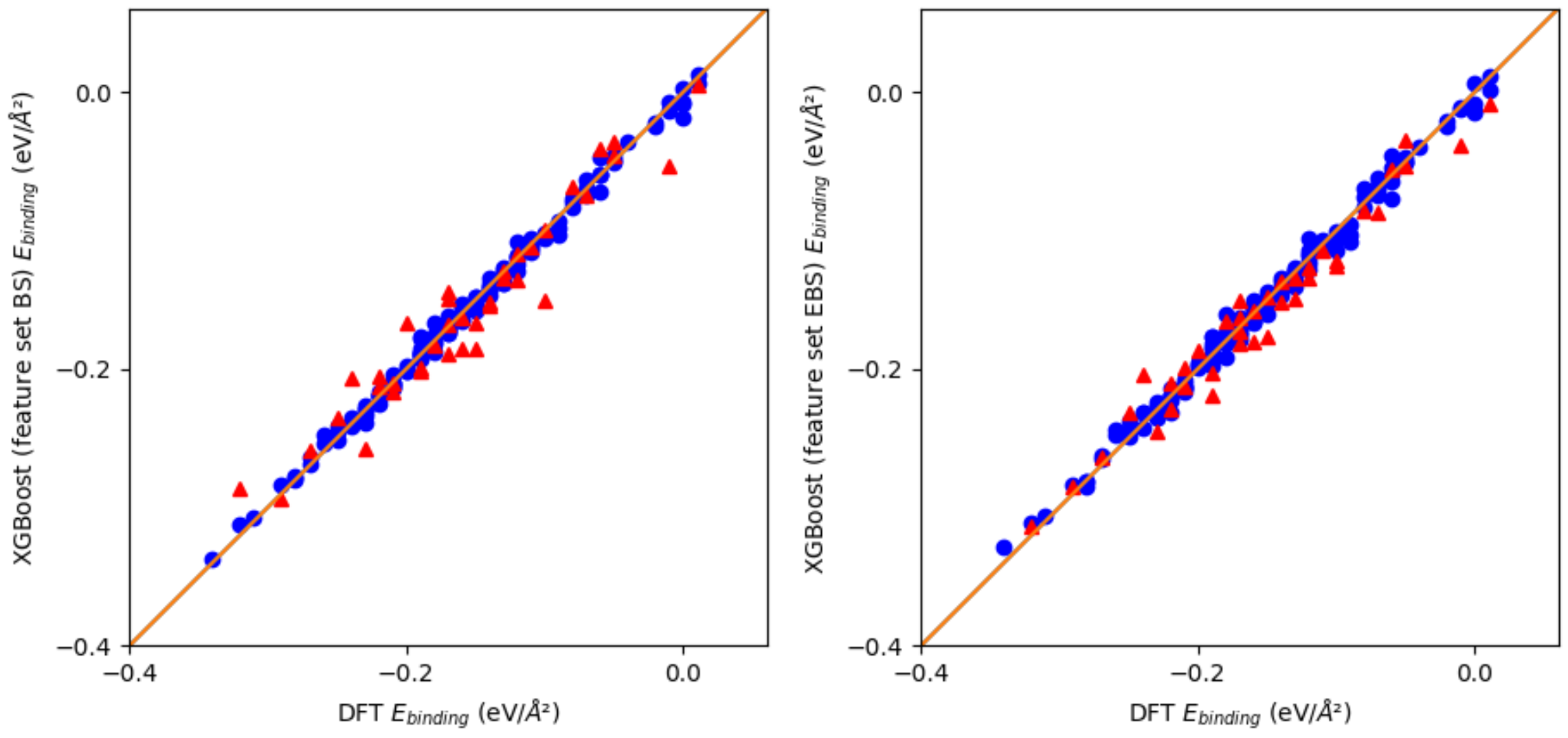

**Figure S3.** ML fitting plots for the 3[rd] ML model set fitting $E_{DFT}$ including $CT_{DFT}$ as one feature.

**Table S5.** ML evaluations for the 4[th] ML model set generating ML-predicted $E_{ML}$ including $CT_{ML}$ as one feature.

| | | Feature set | | | | | | |
|---|---|---|---|---|---|---|---|---|
| | | E | B | S | EB | ES | BS | **EBS** |
| RF | Training $R^2$ | 0.95 | 0.95 | 0.94 | 0.95 | 0.95 | 0.95 | 0.95 |
| | Testing $R^2$ | 0.84 | 0.85 | 0.83 | 0.85 | 0.90 | 0.87 | 0.88 |
| | Training RMSE (eV/Å²) | 0.02 | 0.02 | 0.02 | 0.03 | 0.02 | 0.02 | 0.02 |
| | Testing RMSE (eV/Å²) | 0.03 | 0.03 | 0.03 | 0.03 | 0.02 | 0.03 | 0.03 |
| XGBoost | Training $R^2$ | 0.99 | 0.99 | 0.98 | 0.99 | 0.99 | 0.99 | **0.99** |
| | Testing $R^2$ | 0.88 | 0.92 | 0.89 | 0.91 | 0.93 | 0.95 | **0.96** |
| | Training RMSE (eV/Å²) | 0.01 | 0.01 | 0.01 | 0.01 | 0.01 | 0.01 | **0.01** |
| | Testing RMSE (eV/Å²) | 0.03 | 0.02 | 0.03 | 0.02 | 0.02 | 0.02 | **0.02** |

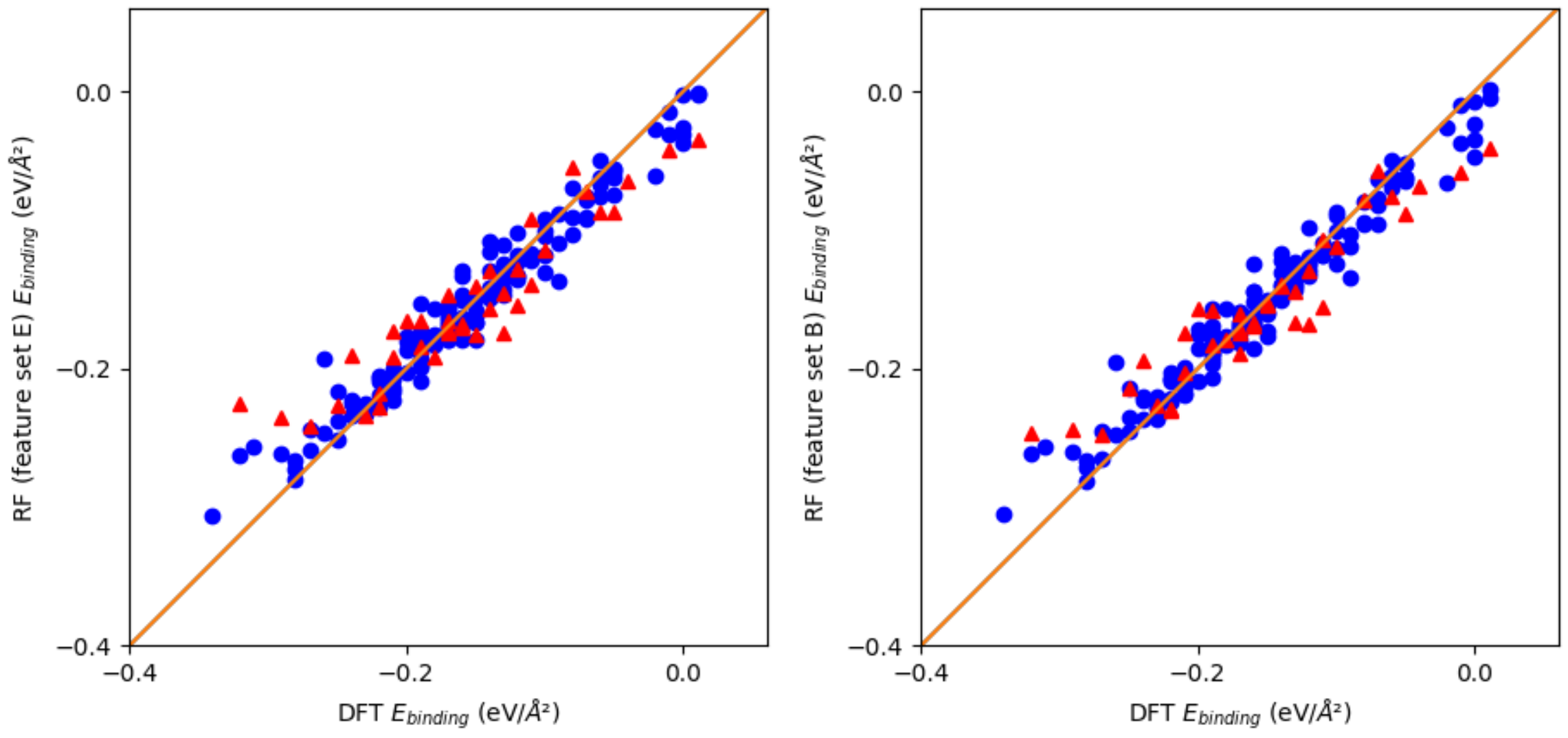

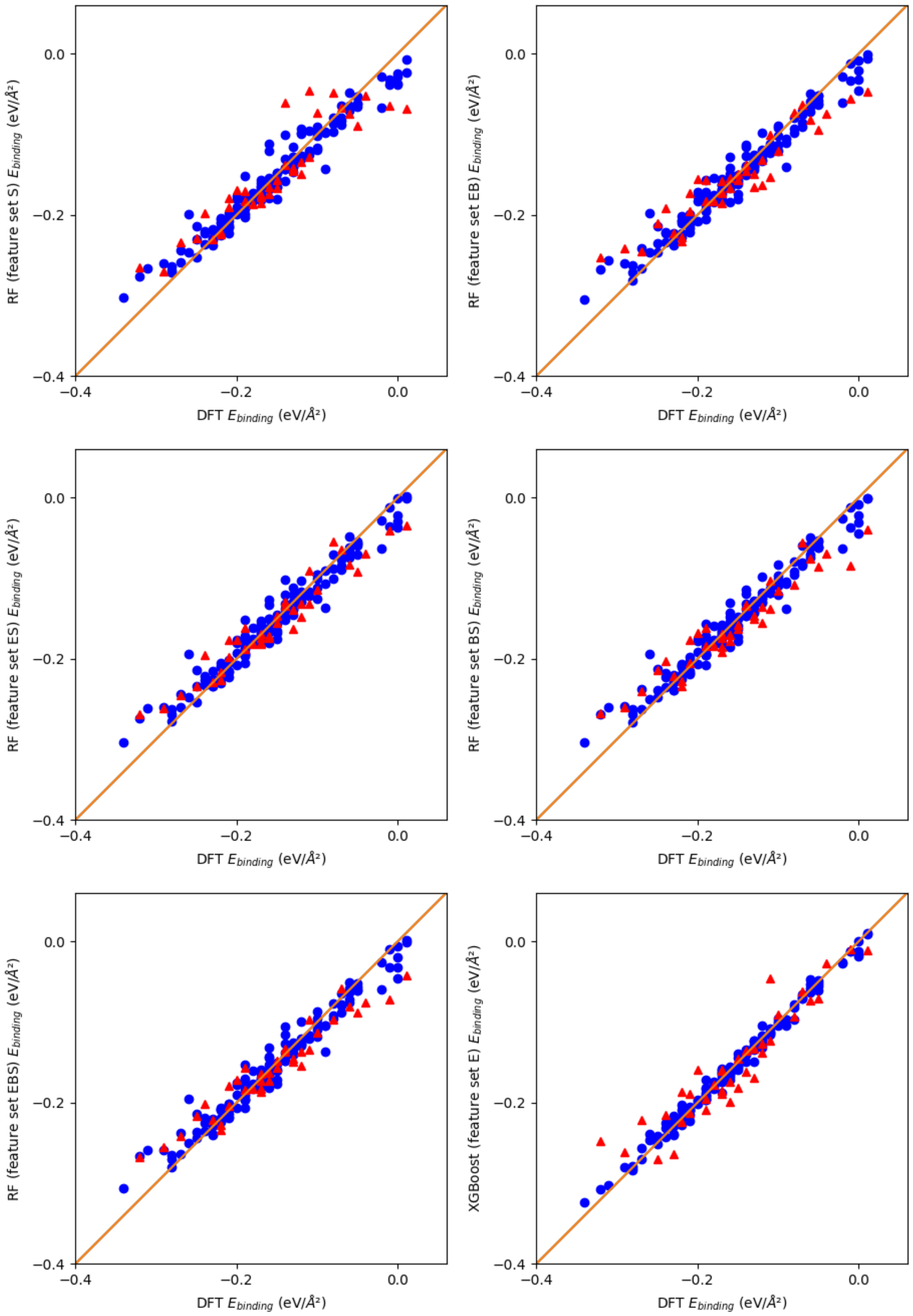
RF (feature set S) $E_{binding}$ (eV/Å²)
RF (feature set EB) $E_{binding}$ (eV/Å²)
RF (feature set ES) $E_{binding}$ (eV/Å²)
RF (feature set BS) $E_{binding}$ (eV/Å²)
RF (feature set EBS) $E_{binding}$ (eV/Å²)
XGBoost (feature set E) $E_{binding}$ (eV/Å²)
DFT $E_{binding}$ (eV/Å²)
0.0
−0.2
−0.4

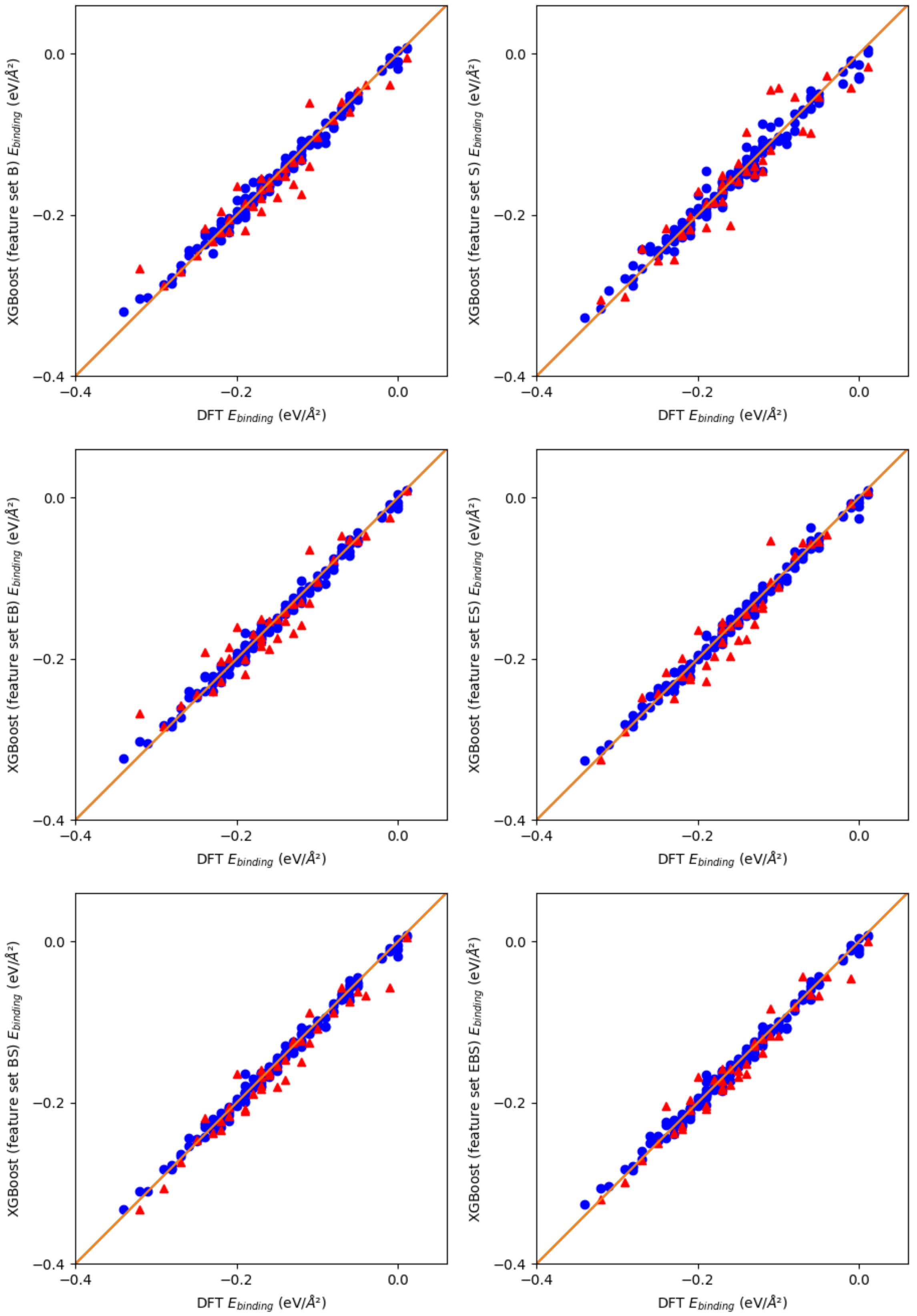

**Figure S4.** ML fitting plots for the 4th ML model set fitting $E_{DFT}$ including $CT_{ML}$ as one feature.

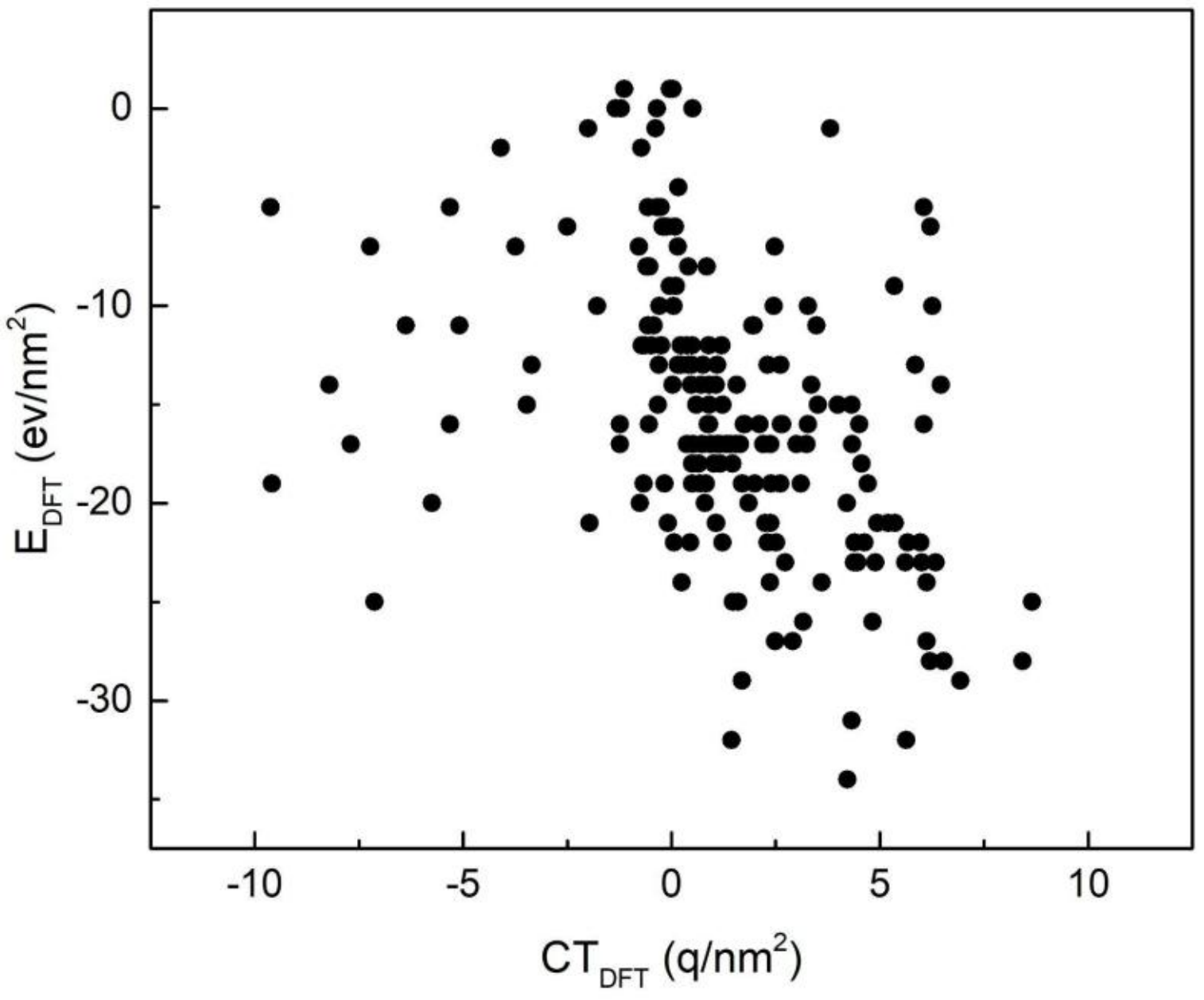

**Figure S5.** The plot of DFT-calculated $E_{binding}$ and charge transfer CT.

## 4. ML screening.

**Table S6.** Potential substrates for borophene growth by ML screening. (* Materials ID means the boride is an existing materials in Materials Project database with a materials ID; A boride without materials ID means it is a hypothetic structure by the substitution of the metal atoms from an existing structure.)

| v1/6 Growth | | | | | |
|---|---|---|---|---|---|
| **No.** | **Formula** | **Material ID *** | **$E_{binding}$ (eV/Å$^2$)** | **Surface Type** | **Bulk Space Group** |
| 1 | AlB2 | | 0.02 | B-rich | 59 |
| 2 | AlB2 | | 0.01 | B-rich | 194 |
| 3 | AlB | | -0.02 | B-rich | 141 |
| 4 | KB | mp-1059907 | -0.02 | Stoichiometry-balanced | 225 |
| 5 | Ag2B | | -0.02 | B-rich | 140 |
| 6 | Ag7B3 | | -0.02 | B-rich | 186 |
| 7 | AlB | | -0.02 | B-rich | 187 |
| 8 | Al7B3 | | -0.02 | B-rich | 186 |
| 9 | Al2B | | -0.02 | B-rich | 140 |
| 10 | RbB | mp-1061023 | -0.02 | Stoichiometry-balanced | 225 |
| 11 | Ag3B | | -0.02 | B-rich | 62 |
| 12 | AlB | | -0.02 | B-rich | 62 |
| 13 | CaB2 | mp-1009695 | -0.02 | B-rich | 191 |
| 14 | AgB2 | | -0.02 | B-rich | 194 |
| 15 | Al3B2 | | -0.02 | B-rich | 127 |
| 16 | AlB | | -0.02 | B-rich | 63 |
| 17 | Al3B | | -0.03 | B-rich | 62 |
| 18 | Al23B50 | mp-530274 | -0.03 | B-rich | 2 |
| 19 | In3B | mp-1184483 | -0.03 | B-rich | 221 |
| 20 | In3B | mp-1184492 | -0.03 | B-rich | 225 |
| 21 | AlB | | -0.03 | B-rich | 194 |
| 22 | Ag3B2 | | -0.03 | B-rich | 127 |
| 23 | TmB2 | mp-800 | -0.03 | B-rich | 191 |
| 24 | AgB | | -0.03 | B-rich | 194 |
| 25 | Na3B | mp-1186020 | -0.03 | B-rich | 225 |
| 26 | Tl3B | mp-1187526 | -0.03 | B-rich | 221 |
| 27 | Tl3B | mp-1187632 | -0.03 | B-rich | 225 |
| 28 | Na3B | mp-1186025 | -0.03 | B-rich | 139 |
| 29 | AgB | | -0.03 | B-rich | 141 |
| 30 | AgB | | -0.03 | B-rich | 187 |
| 31 | NaB | mp-1056860 | -0.03 | B-rich | 225 |
| 32 | AgB | | -0.04 | B-rich | 63 |
| 33 | AgB2 | | -0.04 | B-rich | 59 |
| 34 | AgB | | -0.04 | B-rich | 62 |
| 35 | MgB | | -0.04 | B-rich | 194 |
| 36 | Ag7B3 | | -0.04 | M-rich | 186 |
| 37 | Ag3B | | -0.04 | M-rich | 62 |
| 38 | BeB | | -0.04 | M-rich | 62 |
| 39 | BeB | | -0.04 | M-rich | 63 |
| 40 | Mg2B | | -0.04 | B-rich | 140 |
| 41 | Mg3B2 | | -0.04 | B-rich | 127 |
| 42 | MgB | | -0.04 | B-rich | 63 |
| 43 | MgB | | -0.04 | B-rich | 141 |
| 44 | MgB | | -0.04 | B-rich | 187 |
| 45 | CaB6 | mp-865 | -0.04 | B-rich | 221 |
| 46 | AlB2 | | -0.04 | M-rich | 59 |
| 47 | CsB | mp-1060982 | -0.04 | Stoichiometry-balanced | 225 |
| 48 | Ag2B | | -0.04 | M-rich | 140 |
| 49 | CaB4 | mp-1213975 | -0.04 | B-rich | 127 |
| 50 | Al23B50 | mp-530274 | -0.04 | M-rich | 2 |
| 51 | TmB4 | mp-568143 | -0.05 | B-rich | 127 |
| 52 | TmB12 | mp-1104905 | -0.05 | B-rich | 225 |
| 53 | TmB6 | mp-571352 | -0.05 | B-rich | 221 |
| 54 | AlB | | -0.05 | B-rich | 225 |
| 55 | RbB | mp-1061023 | -0.05 | B-rich | 225 |
| 56 | Sn3B | mp-978855 | -0.05 | M-rich | 221 |
| 57 | KB | mp-1059907 | -0.05 | B-rich | 225 |
| 58 | Cd3B | mp-1183580 | -0.05 | B-rich | 139 |
| 59 | AlB | | -0.05 | M-rich | 62 |
| 60 | AlB | | -0.05 | M-rich | 63 |

| **v1/9 Growth** | | | | | |
|---|---|---|---|---|---|
| **No.** | **Formula** | **Material ID *** | **$E_{binding}$ (eV/Å$^2$)** | **Surface Type** | **Bulk Space Group** |
| 1 | AlB2 | | 0.01 | B-rich | 59 |
| 2 | AlB2 | | 0.01 | B-rich | 194 |
| 3 | AlB | | -0.01 | B-rich | 63 |
| 4 | AlB | | -0.01 | B-rich | 62 |
| 5 | Al3B2 | | -0.01 | B-rich | 127 |
| 6 | Al7B3 | | -0.01 | B-rich | 186 |
| 7 | AlB | | -0.01 | B-rich | 141 |
| 8 | AlB | | -0.01 | B-rich | 187 |
| 9 | Al2B | | -0.01 | B-rich | 140 |
| 10 | Al3B | | -0.01 | B-rich | 62 |
| 11 | BeB | | -0.02 | M-rich | 62 |
| 12 | Ag7B3 | | -0.02 | B-rich | 186 |
| 13 | AgB2 | | -0.02 | B-rich | 194 |
| 14 | CuB5 | mp-1181765 | -0.02 | M-rich | 61 |
| 15 | AlB | | -0.02 | B-rich | 194 |
| 16 | CaB2 | mp-1009695 | -0.02 | B-rich | 191 |
| 17 | Ag2B | | -0.02 | B-rich | 140 |
| 18 | AlB2 | | -0.03 | M-rich | 59 |
| 19 | Al23B50 | mp-530274 | -0.03 | B-rich | 2 |
| 20 | In3B | mp-1184483 | -0.03 | B-rich | 221 |
| 21 | In3B | mp-1184492 | -0.03 | B-rich | 225 |
| 22 | AgB2 | | -0.03 | B-rich | 59 |
| 23 | TmB2 | mp-800 | -0.03 | B-rich | 191 |
| 24 | Ag3B | | -0.03 | B-rich | 62 |
| 25 | Al23B50 | mp-530274 | -0.03 | M-rich | 2 |
| 26 | Tl3B | mp-1187526 | -0.03 | B-rich | 221 |
| 27 | Tl3B | mp-1187632 | -0.03 | B-rich | 225 |
| 28 | AlB2 | | -0.03 | M-rich | 194 |
| 29 | Na3B | mp-1186020 | -0.03 | B-rich | 225 |
| 30 | Na3B | mp-1186025 | -0.03 | B-rich | 139 |
| 31 | Be4B25 | mp-1227634 | -0.03 | M-rich | 13 |
| 32 | NaB | mp-1056860 | -0.04 | B-rich | 225 |
| 33 | KB | mp-1059907 | -0.04 | Stoichiometry-balanced | 225 |
| 34 | Mg2B | | -0.04 | B-rich | 140 |
| 35 | Mg3B2 | | -0.04 | B-rich | 127 |
| 36 | RbB | mp-1061023 | -0.04 | Stoichiometry-balanced | 225 |
| 37 | CaB4 | mp-1213975 | -0.04 | B-rich | 127 |
| 38 | Ag7B3 | | -0.04 | M-rich | 186 |
| 39 | Ag3B | | -0.04 | M-rich | 62 |
| 40 | W2B5 | mp-8079 | -0.04 | M-rich | 166 |
| 41 | WB2 | | -0.04 | M-rich | 59 |
| 42 | BeB | | -0.04 | Stoichiometry-balanced | 62 |
| 43 | CaB6 | mp-865 | -0.04 | B-rich | 221 |
| 44 | CaB4 | mp-1213975 | -0.04 | M-rich | 127 |
| 45 | AlB | | -0.04 | B-rich | 225 |
| 46 | Be3B2 | | -0.04 | M-rich | 127 |
| 47 | Ag2B | | -0.04 | M-rich | 140 |
| 48 | BeB | | -0.05 | M-rich | 63 |
| 49 | BeB2 | | -0.05 | M-rich | 59 |
| 50 | BeB | | -0.05 | Stoichiometry-balanced | 141 |
| 51 | BeB | | -0.05 | Stoichiometry-balanced | 187 |
| 52 | TmB4 | mp-568143 | -0.05 | B-rich | 127 |
| 53 | BeB | MB | -0.05 | M-rich | 141 |
| 54 | BeB | MB | -0.05 | M-rich | 187 |
| 55 | TmB12 | mp-1104905 | -0.05 | B-rich | 225 |
| 56 | TmB6 | mp-571352 | -0.05 | B-rich | 221 |
| 57 | BeB | | -0.05 | Stoichiometry-balanced | 63 |
| 58 | AlB | | -0.05 | M-rich | 62 |
| 59 | AlB | | -0.05 | M-rich | 63 |
| 60 | AlB | | -0.05 | M-rich | 141 |
| 61 | Be7B3 | | -0.05 | M-rich | 186 |
| 62 | Sn3B | mp-978855 | -0.05 | M-rich | 221 |
| 63 | Cd3B | mp-1183580 | -0.05 | B-rich | 139 |
| 64 | RbB | mp-1061023 | -0.05 | B-rich | 225 |
| 65 | KB | mp-1059907 | -0.05 | B-rich | 225 |